# Complement or substitute? How AI increases the demand for human skills

Elina Mäkelä[a], Matthew Bone[ab], Farah Nanji[a], Mareike Sehrer[a], Fabian Stephany[acd]

a) Oxford Internet Institute, University of Oxford, UK, b) Burning Glass Institute, USA, c) Institute for New Economic Thinking, Oxford Martin School, d) Bruegel, Brussels, Belgium.

**Email:** fabian.stephany@oii.ox.ac.uk

## Abstract

Artificial Intelligence (AI) is transforming the nature of work, yet there is limited empirical evidence on how it affects demand for human skills. This paper examines whether AI adoption increases the prevalence and value of human capabilities that complement technical AI skills, such as analytical thinking, resilience, or ethical judgment, within and beyond AI-intensive job roles. Using a dataset of nearly 30 million job postings from the US, the UK and Australia, between 2018 and 2024, we distinguish between internal effects (within AI roles) and external effects (in non-AI roles) across companies, industries, and regions. This paper has three main findings. First, we find that AI-intensive roles are significantly more likely to require complementary non-technical capabilities, such as analytical thinking, resilience, and digital literacy. Second, these complementary skills are associated with meaningful wage premiums, particularly in managerial, sales or finance roles working with AI. Third, we show that AI diffusion has potential spillover effects: as AI adoption rises within companies, industries, and regions, demand for complementary skills increases even in non-AI roles while demand for substitutable skills – summarisation, translation or customer service – decreases. These trends hold across geographies, including the United States, United Kingdom, and Australia, confirming the robustness of our findings. Together, these findings indicate that AI is not simply replacing tasks or requiring more AI developer skills, it may be transforming workforce skill requirements to favor human attributes that enhance collaboration with intelligent systems.

## Significance Statement

Using a dataset of about 30 million job postings between 2018 and 2024 from the US, the UK, and Australia, we provide correlative evidence that AI does not simply replace jobs, it reshapes them by elevating the demand and monetary value of complementary, human skills such as analytical thinking, resilience, and ethical judgment. These effects are not confined to technical roles: as AI diffuses across companies, demand rises for human capabilities in adjacent non-AI roles across industries and regions in all three countries. Our findings offer timely insights for workforce development, organizational planning, and policy design, underscoring the need to invest in the human skills that enable effective collaboration with artificial intelligence systems.

## Introduction

The effects of Artificial Intelligence (AI) on the demand for human skills is a topic of growing significance, with critical implications for policymakers and economists (1, 2, 3, 4, 5, 6). While much of prior work has focused on substitution, where AI automates routine manual or cognitive tasks (4, 7), more recent work has begun to examine how human-AI interactions augments work (3, 8, 9, 10). The increasing demand and financial rewards for AI developer skills are well researched (11, 12) in light of the recent surge in AI adoption (13, 14), but the specific ways in which AI adoption reshapes demand for non-technical, human-centered capabilities remain poorly understood (9, 15). These capabilities are often cited as critical for working alongside AI systems (11, 12), yet there is limited consensus on which specific skills are most affected, how their demand evolves, or whether they are rewarded in the labor market (16). This research centers on a key empirical question: whether complementary skills, such as digital literacy, teamwork, or analytical thinking, become more prevalent and valuable as AI adoption grows.

Specifically, we investigate whether growing demand for AI-centric work not only reshapes skill requirements within AI-intensive roles but also exerts broader influence across companies, industries, and regions. Much of the debate around AI and work focuses on substitution or augmentation effects and looks at technical skills like machine learning and data analysis (17, 18). This research addresses a gap by focusing on the emerging human-centered skill demand, beyond technical AI skills, offering practical insights for policymakers, employers, and job-seekers. *SI Appendix*, section S2 details how AI roles and skill categories were defined and classified.

We address this gap by empirically distinguishing between two types of effects, **internal** and **external**, that arise from the diffusion of artificial intelligence and reflect heterogeneous changes in workers' task composition. This distinction builds on previous work emphasizing that new technologies can differentially reshape the combination of tasks and skills required within and across occupations (1,2,4,7). **Internal effects** refer to shifts in skill requirements *within* roles that are directly associated with AI-related work. However, we argue that these effects are unlikely to be uniform across AI-related occupations. For workers who have long been engaged in the development and deployment of AI technologies, such as data scientists and machine learning engineers, the advent of generative AI may not substantially alter underlying task compositions (AI-creators). By contrast, occupations in managerial, business and finance, sales, and related domains have historically been less directly exposed to AI technologies but are now increasingly integrating AI into core work processes (AI-users). For these workers, the increase in human-AI interaction is more likely to induce meaningful changes in task composition, increasing the value of complementary human-centric skills such as judgment, ethical reasoning, adaptability, and analytical thinking. In this sense, AI-related technical capabilities may amplify complementarities between technical inputs and distinctly human social and cognitive capabilities (19). **External effects** capture broader changes that occur as AI demand differs across firms, industries, and regions, affecting workers who may not directly interact with AI systems themselves. Organizational adoption of AI can reshape workflows, alter team structures, introduce new technical interfaces, and raise coordination and oversight requirements, thereby shifting skill demand throughout the workforce. Prior research shows that such organizational adjustments often increase demand for interpersonal, problem-solving, and ethical judgement skills, even outside explicitly technical roles (1,6). These external effects thus reflect indirect channels through which AI adoption influences labor demand beyond the set of workers formally classified as working in AI-related roles.

We formalize this framework into two sets of hypotheses:

Internal effects (within AI roles):

H1a: AI roles are more likely to require complementary skills.
H1b: Complementary skills in AI roles are associated with higher wages.

External effects (beyond AI roles):

H2a: In companies, industries, or regions with high AI adoption, non-AI roles are more likely to require complementary skills.
H2b: In those same contexts, non-AI roles are less likely to require substitutable skills.

To investigate these patterns, we analyze a large dataset of approximately 30 million online job postings from the US, the UK, Australia and New Zealand, between 2018 and 2024. We first reveal a substantial increase in the demand for skills seen as human-centric and complementary to AI, such as analytical thinking, teamwork, and resilience, when comparing AI roles to non-AI roles. Along with increases in demand, we find a growing wage premium for these skills within AI-focused roles in domains such as management, sales, finance, or admin support. We then delve deeper into complementary skills, demonstrating that demand and wage premiums are particularly pronounced for some skills, and that specific combinations of AI and complementary skills appear especially popular and profitable within AI roles.

Finally, we then turn to external effects, conducting company, industry, and region-level analyses. We find that the diffusion of AI is associated with rising demand for complementary skills and falling demand for some substitutable skills, even in non-AI roles. We find these results to be robust across the US, the UK, Australia and New Zealand. This study sheds a new light on how AI adoption is reshaping skill demands across the economy. By identifying which skills gain value in an AI-integrated labor market, our findings offer practical guidance for policymakers, employers, and workers navigating the future of work.

## Methods: Data, Skill Classification, and AI Roles Identification

A significant challenge in measuring the effects of AI on the future of work is the lack of high quality, empirically grounded data (6). Following the literature, we assume that employers' online vacancy postings reflect evolving employer demand and valuation for different skill sets (1, 19). The analysis of "internal effects" is based on a sample of 10 million online job postings (n = 9,998,342) from the United States between January 2018 and December 2024. The data, created by Lightcast and provided by the Burning Glass Institute (BGI), were scraped from over 65,000 sources. Each vacancy was cleaned and labeled with a job title, occupation category, salary, and a detailed skill list using Lightcast's proprietary methods. This data has been previously used to study the labor market effects of technological change (1, 11, 20, 21, 22, 23) and closely align with trends reported by the Bureau of Labor Statistics (BLS) (24) and JOLTS, providing the “near-universe” of online vacancy data (1). For the analysis of “external effects”, the sample is extended to include job postings from the United Kingdom as well as Australia and New Zealand between 2018 and 2024, yielding 30 million job postings in total. Details on the data sources and sample construction are outlined in the appendix, Tables S1 and S2.

To distinguish between AI and non-AI roles, we use Lightcast’s skill subcategory “Artificial Intelligence and Machine Learning (AI/ML)" to define a set of AI skills. Any job posting that lists at least one of these skills is classified as an AI role. This skills-based approach (2, 11, 20) is typically used for identifying AI roles in job vacancy data (1,20). We further identify AI-complementary and AI-substitutable skills based on a synthesis of prior research on automation, skill-based technological change, and labor economics (2, 4, 6, 28). This review yields a set of seven ‘complementary’ and four ‘substitutable’ skills clusters. A job posting was coded as demanding complementary or substitutable skills if it lists at least one skill from the corresponding cluster. Section S2 of the supplementary appendix provides a detailed methodology and complete list of skills.

## Analysis of Internal Effects

To begin our examination on how skill demand and salaries have evolved in AI vs. non-AI roles, we inspect developments over time from 2018 to 2024. We calculated the annual share of job postings requiring complementary or substitutable skills and plotted the differences in AI vs. non-AI roles. Odds ratios were computed as the probability of a given skill appearing in AI roles relative to non-AI roles, with values above 1 indicating higher demand in AI jobs. To visualize trends, we plotted the odds over time for each skill cluster.

*(FIGURE 1 & CAPTION)*

A general trend of increasing prominence of complementary skills both in AI-related and non AI-related roles can be observed across the time period from 2018 to 2024 (lower chart). Five of the seven skill clusters - analytical thinking, technical proficiency, self-efficiency, working with others (teamwork), and resilience - consistently appear more often in AI-related postings, with resilience exhibiting the sharpest increase over time. Ethics skills on the other hand, only appear more frequently in AI-related roles in the later years, from 2022, coinciding with the launch of the first mass-market generative AI chatbot ChatGPT. Meanwhile, digital literacy skills consistently appear more frequently in non AI-related roles.

We further examine the demand for complementary and substitutable skills in job postings through regression analysis. Specifically, we employ a logistic regression to assess how the demand for different skills increases in AI-related job postings compared to non-AI related postings. The regression controls for a large set of potentially influential features, such as required experience, education, or industry. The upper panel of Figure 2 plots the coefficients of interest and captures a significant positive demand effect for five of seven complementary skills, in accordance with our hypothesis H1a. Reflecting the observations made from Figure 1, no significant effect can be detected for “Ethics” and “Digital Literacy” skills. The specification, list of control variables, and regression tables of this analysis are included in the *SI Appendix*, Section 3 and Table S8.

Within AI-related roles, the extent to which recent advances in generative artificial intelligence reshape skill demand is likely to vary across occupations. For "AI-creator" occupations that have long been closely involved in the development of AI technologies, such as data scientists and machine learning engineers, the introduction of generative AI may not fundamentally alter underlying skill requirements. In contrast, "AI-user" occupations in which AI technologies have historically played a more limited role, particularly managerial, business and finance, sales and related, and office and administrative support roles, are increasingly integrating AI into core work processes. We therefore expect the complementarity between AI-related skills and human-centric skills to be most pronounced in these occupations.

The rightmost panel of Figure 2 focuses on these roles, which have previously been less affected by AI but are now experiencing more substantial changes in skill demand. Specifically, we isolate management, business and finance, sales and related, as well as office and admin support occupations. "Self-efficiency" skills exhibit significant positive demand effects across all four occupation groups, with the coefficients being particularly pronounced in sales, and office and administration occupations. In the sample of management, and business and finance jobs, significant positive demand effects are captured for the majority of complementary skills, with the exception of "ethics" and "digital literacy". The coefficients for these four selected occupations are generally much larger than those found for the overall sample.

## Valuation of Skills

Regression as a method of analysis has already been used to understand the relationships between artificial intelligence and patterns in demand and wages in the labor market (1, 12). To assess the wage reward for complementary skills, we implement a linear regression comparing log-transformed posted salaries in AI roles demanding complementary or substitutable skills with AI roles that do not demand the given skills. The left panel of Figure 2 further shows that the increased demand for complementary skills in AI roles is backed up with statistically significant salary premia. Specifically, "Resilience" commands a 5% salary premium. However, no salary premia can be confidently detected for the other complementary skills, providing only limited evidence in favour of hypothesis H1b. Due to sample size restrictions, we refrain from performing the same wage regression analysis for the individual "AI-user" occupations. To nevertheless provide descriptive insights into potential skill-related salary effects, the lower panel of Figure 2 displays two histograms of salary distributions within Management, and Business and Finance roles. In both occupations, roles requiring "Resilience" and "Working with others" skills respectively, are associated with higher salaries, compared to other AI-related roles that do not require the corresponding skill. While this does not offer any causal evidence of complementary skills yielding wage premia, the analysis does provide descriptive evidence that the interaction between AI technologies and human-specific social and cognitive skills is valued in the workplace. The specification, list of control variables, and regression tables of this analysis are included in the *SI Appendix*, Section S3 and Table S12.

*(FIGURE 2 & CAPTION)*

## External Effects: Companies, Industry, and Region across the US, UK, Australia and New Zealand

We extend our analysis beyond AI-related roles to assess how AI adoption reshapes skill demand across the broader economy. We test whether, as AI penetrates workflows within companies, industries, or regions, its effect on skill requirements extends even to roles that do not directly work with AI. We hypothesise that rising AI adoption (a higher prevalence of AI roles) within a company, industry, and region increases the demand for complementary skills and decreases the need for substitutable skills, even in non-AI positions. For example, regulatory authorities in regions deploying autonomous systems like self-driving cars may increasingly require ethics-related skills, and HR recruiters may need stronger digital literacy as AI becomes more embedded in recruitment processes, despite not working directly with AI systems themselves.

The upper panel of Figure 3 plots the relationship between the number of AI roles in a US company against demand for complementary and substitutable skills in non-AI roles. Each data point represents a company in the year 2024, with trend lines fitted using linear regression and estimated coefficients shown directly on the plots; see complete set of results in *SI Appendix*, Section S4.13. A positive association is observed for complementary skills ($\beta = 0.075$, $p<0.01$), and a corresponding negative association for substitutable skills ($\beta = -0.09$, $p<0.01$), offering a visual and statistical representation of these opposing trends.

*(FIGURE 3 & CAPTION)*

The lower panel of Figure 3 captures the external demand effect at the company, industry, and region level, for the United States, United Kingdom, Australia and New Zealand. Each plot displays the calculated marginal effects from the corresponding fractional logistic regression analyses conducted on the sample from 2018 to 2024, including various controls. In support of H2a, AI prevalence is associated with increased demand for complementary skills in non-AI roles across all three levels and countries. At the industry level, as well as the region level in Australia and New Zealand specifically, the marginal effects are notably larger in magnitude than at the company level. Conversely, marginal effects for substitutable skills in non-AI roles are consistently negative in the United States across all levels, and either positive or not significantly different from zero for the United Kingdom and Australia and New Zealand, providing favourable evidence for hypothesis H2b.

The opposing demand effects for complementary and substitutable skills appear most clearly at the firm-level. Within a firm, AI technologies might reach a wider sphere of influence as workflows are likely more interconnected than within industries and regions. However, the upper panel of Figure 3 also highlights high heterogeneity of firms in terms of AI adoption and skills requirements. It's possible that this correlation is simply capturing the heterogeneity of firms that require different bundles of skills and technology, rather than measuring the impacts of AI adoption over time.

We further explore how each subcategory of complementary and substitutable skills contributes to these observed effects in Figure S11 of the *SI Appendix*. Most notably, we find the decreased demand for substitutable skills to be largely driven by lower demand for "customer service" skills. We consistently observe this pattern for the company, industry, and region-level analyses. It should further be noted that the wider impact of AI technologies on workflows and the prioritization of skills is likely heterogeneous across firms.

Figure 3 provides indications that the spread of AI may have ripple effects beyond roles working directly with AI technologies. As AI-related roles increase in a company, industry, or region, the workflows of adjacent non-AI roles are also redefined, favoring skills that complement rather than compete with AI systems. This indicates that AI's influence on jobs is systemic and rooted in human-AI interactions and interdependencies, reinforcing the case for broad-based skill adaptation focused on human–AI complementarity

To our knowledge, these findings are among the first to quantify the external spillovers in skill demand linked to AI adoption. They reveal how workplaces are subtly but significantly reconfigured as demand shifts toward complementary human capabilities, challenging purely substitution-based narratives and highlighting the nuanced impact of AI on the broader workforce. These findings carry several important implications. For workers, they underscore the value of acquiring and developing AI-complementary skills, such as analytical thinking, resilience, and working with others, to thrive in AI-integrated environments. For companies, they emphasize the need for proactive workforce planning and reskilling to meet these evolving demands and benefit from emerging human-AI synergies. And for policymakers, the regional spillover effects highlight the importance of targeted education and workforce development programs in high-AI-adoption areas, alongside safety nets to support workers displaced by automation.

## Conclusion

This study provides new empirical evidence on how AI adoption is shaping the demand for complementary skills in the labor market. Moving beyond the conventional "substitution" narrative, we trace the growing presence and valuation of human-AI interaction, affecting both roles working directly with AI technologies, as well as non AI-related workflows within environments of high AI prevalence. Technical skills that develop AI remain highly valuable, as prior research shows, but the potential of human skills like analytical thinking, resilience, and technical proficiency, when paired with AI technologies, is reflected by their increasing demand in AI jobs and, in some cases, higher wages. This effect is particularly pronounced in occupations likely to engage with AI technologies as users rather than creators. Our findings further suggest that AI diffusion has measurable spillover effects. As AI penetrates the workflows of more companies, industries, and regions, non-AI roles are redefined to prioritize complementary skills and, in some cases, de-emphasize substitutable ones. These external effects are documented across geographies and persist across companies, industries, and regions.

Overall, our findings suggest that AI is not simply replacing workers. It is transforming the broader ecosystem of work by elevating the importance of human attributes that align with, rather than compete against, AI. For policymakers, the findings highlight the urgency of investing in training programs that foster adaptable, interdisciplinary, and ethically-grounded skill sets. For employers and educators, the results underscore the need to rethink talent development strategies in ways that support human–AI interaction. And for workers, future-proofing their careers increasingly depends on cultivating skills that amplify the unique value of human judgment, analytical thinking, and adaptability in an AI-integrated world.

## Acknowledgments

We thank participants at presentations and seminars at the Oxford Internet Institute, the Brown Bag seminar, the Econ Austria Research Institute, and the Complexity Science Hub Vienna for valuable discussions and feedback that substantially improved this work. We are also grateful to David Sutcliffe for his excellent support in editing the manuscript.

# Figures and Tables

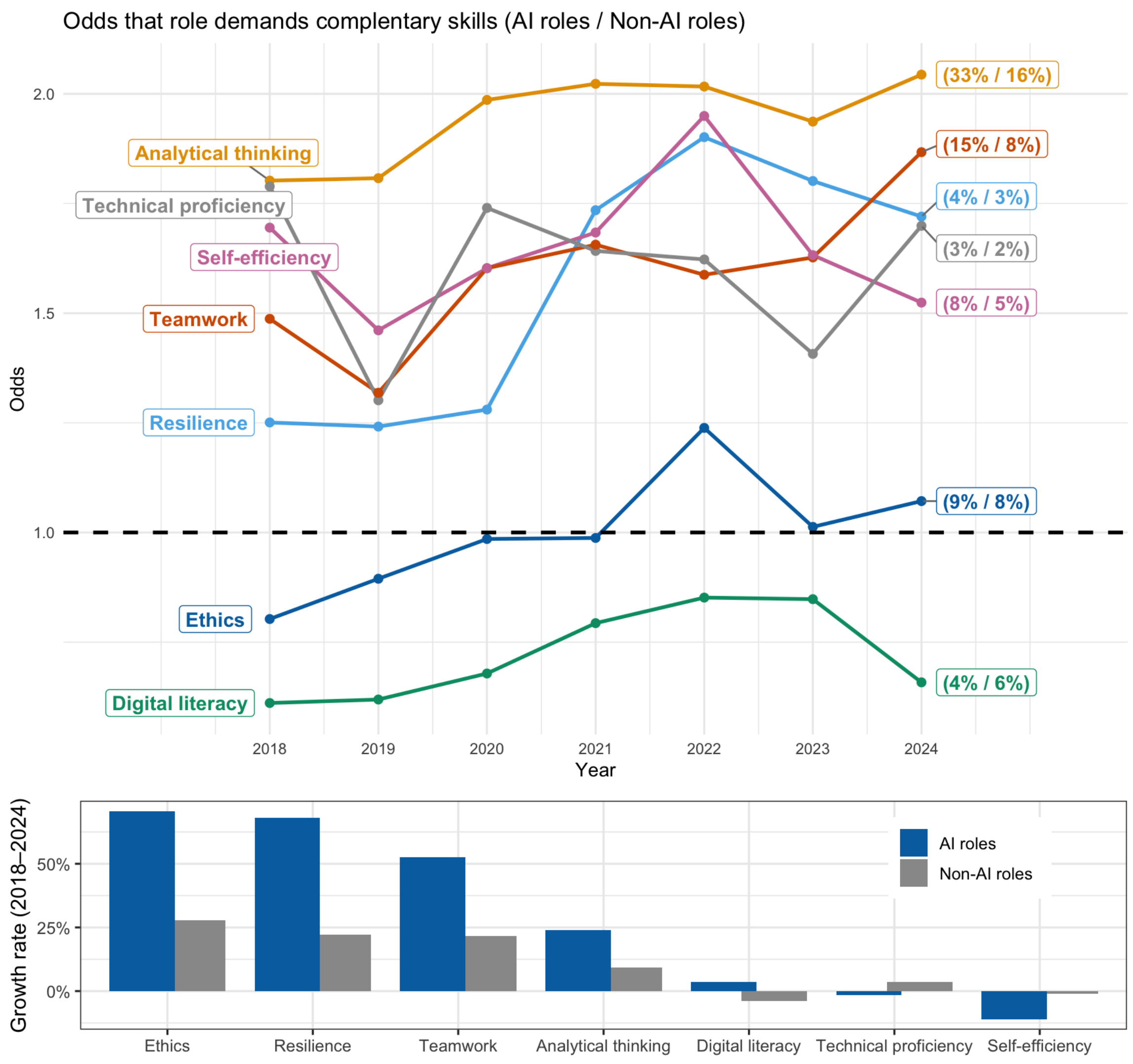

**Fig. 1. Demand for complementary skills, including ethics, resilience, teamwork, analytical thinking, and digital literacy, increased over the observation period for both AI and non-AI roles (lower panel). Across nearly all complementary skills, the likelihood of these skills being requested is significantly higher in AI-related roles than in non-AI roles, as shown in the upper panel.**

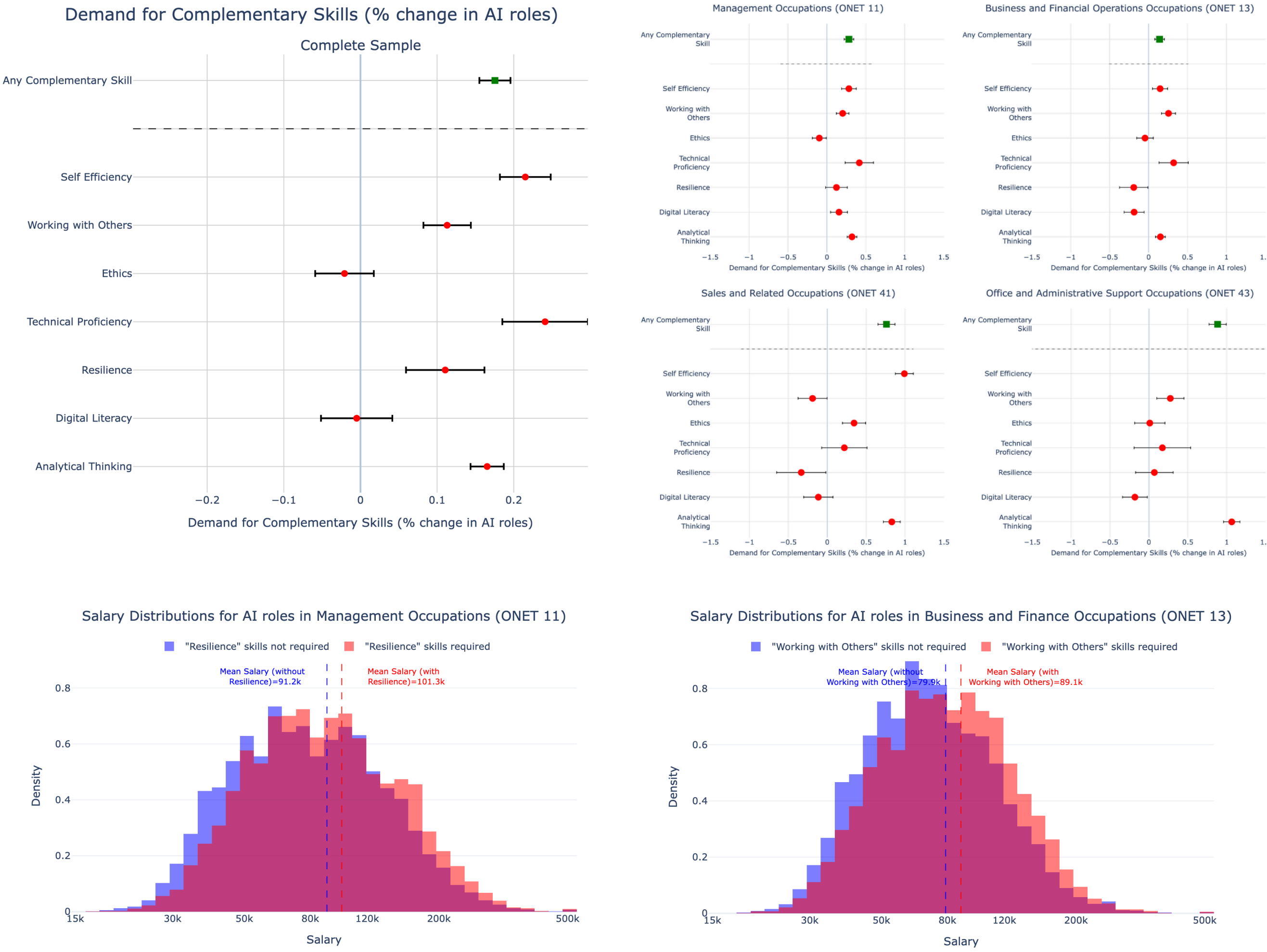

**Fig. 2. The main coefficients of interest of the internal effects regressions (see Appendix Sections S3A and S4B.) The upper left panel shows a positive demand effect for complementary skills in AI roles, except for ethics and digital literacy. A 5% wage premium is detected for resilience in AI roles. The upper right panel decomposes the demand effects for occupations acting as AI-users, again finding positive effects across most complementary skills. The lower panel compares the difference in salary distributions when management roles require resilience skills, and when business and finance roles require working with others skills, demonstrating an upward shift in salaries when roles require the given complementary skill.**

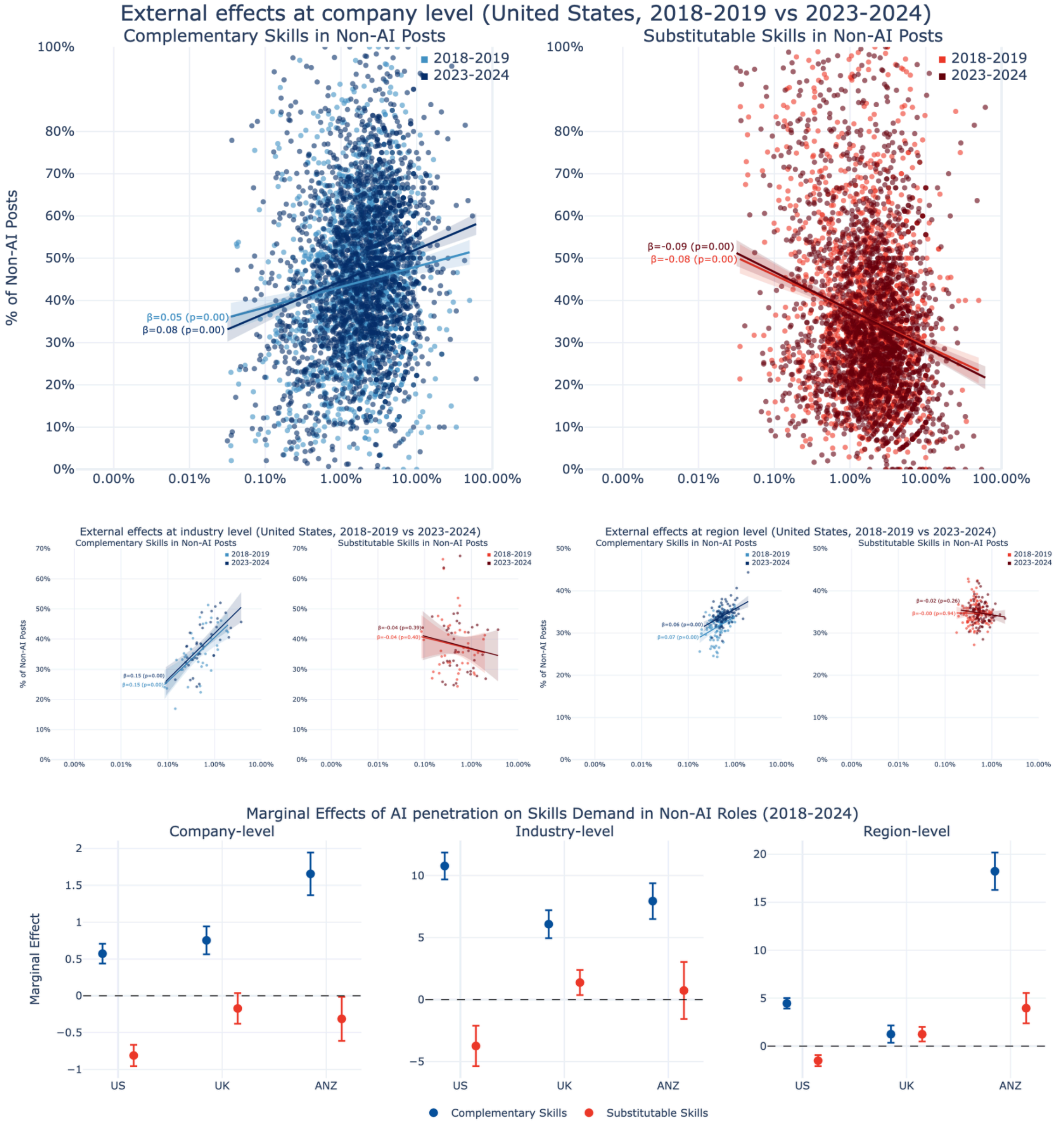

**Fig. 3. External effects of AI penetration on demand for skills in non AI-related roles. The upper panel plots demand for complementary and substitutable skills in non AI-related roles in US companies in 2024. On the left, higher AI penetration within a company is associated with reduced demand for substitutable skills in non-AI roles. On the right, higher AI penetration correlates with increased demand for complementary skills in non-AI roles. The lower panel shows the calculated marginal effects for skills in non AI-roles at the company, industry and region levels for the US, UK, and Australia and New Zealand between 2018 and 2024. Significant positive effects are found for complementary skills across all countries and levels of measurement.**

**Supporting Information for** *Complement or Substitute? How AI Increases the Demand for Human Skills*

**Authors:**

**This PDF file includes:**
Supporting text
Figures S1 - S11
Tables S1 - S14
SI References

This supplementary appendix provides additional detail on the data, variables, models, and robustness checks used in the analysis.

## S1. The Data

Following prior work, we assume that employers' online vacancy postings reflect evolving skill demands and valuations (1, 2). We leverage a dataset of approximately 10 million (n=9,998,342) U.S. online job postings from January 2018 to December 2023, created by Lightcast and provided by the Burning Glass Institute (BGI). These postings are scraped from over 65,000 websites—including job boards, company career pages, and aggregators (Lightcast Data, n.d.)—capturing the "near-universe" of online vacancies (1). Each posting contains information such as job title, occupation category, salary, and a detailed skill list derived from the text.

Lightcast data have previously facilitated studies on technology's labour market effects (e.g., 1, 3, 4, 5, 6, 7). Extensive tests confirm that Lightcast data mirror U.S. vacancy trends from the Job Openings and Labor Turnover Survey (JOLTS) and the Bureau of Labor Statistics (BLS), and the dataset aligns closely with state-level BLS statistics (Tables S1–S2). Although there is some overrepresentation of professional and technical occupations relative to blue-collar jobs—an established characteristic of online postings data (8) — our focus on knowledge work and comparative skill shares minimises any potential biases.

| State | OJV Data | BLS Data |
|---|---|---|
| Alabama | 1.3% | 1.5% |
| Alaska | 0.3% | 0.3% |
| Arizona | 2.7 % | 2.1% |
| Arkansas | 0.6% | 0.9% |
| California | 12.9% | 9.4% |
| Colorado | 2.8 % | 2.4% |
| Connecticut | 1.1% | 1.0% |
| Delaware | 0.3% | 0.3% |
| Florida | 6.2% | 6.5% |
| Georgia | 3.2% | 3.7% |
| Hawaii | 0.4% | 0.3% |
| Idaho | 0.6% | 0.6% |
| Illinois | 4.0% | 4.2% |
| Indiana | 2.0% | 1.8% |
| Iowa | 1.0% | 0.9% |
| Kansas | 1.0% | 0.8% |
| Kentucky | 1.1% | 1.3% |
| Louisiana | 1.0% | 1.5% |
| Maine | 0.3% | 0.5% |
| Maryland | 1.8% | 2.0% |
| Massachusetts | 3.0% | 2.7% |
| Michigan | 2.9% | 2.7% |
| Minnesota | 2.0% | 2.1% |
| Mississippi | 0.5% | 0.9% |
| Missouri | 1.9% | 1.9% |
| Montana | 0.3% | 0.4% |
| Nebraska | 0.7% | 0.6% |
| Nevada | 1.1% | 1.0% |
| New Hampshire | 0.5% | 0.5% |
| New Jersey | 2.5% | 2.4% |
| New Mexico | 0.6% | 0.6% |
| New York | 4.5% | 4.8% |
| North Carolina | 3.3% | 3.6% |
| North Dakota | 0.2% | 0.3% |
| Ohio | 3.6% | 3.6% |
| Oklahoma | 1.1% | 1.3% |
| Oregon | 1.5% | 1.3% |
| Pennsylvania | 3.4% | 4.0% |
| Rhode Island | 0.4% | 0.3% |
| South Carolina | 1.2% | 1.9% |
| South Dakota | 0.3% | 0.3% |
| Tennessee | 2.0% | 2.5% |
| Texas | 8.6% | 8.6% |
| Utah | 1.0% | 1.0% |
| Vermont | 0.2% | 0.2% |
| Virginia | 3.0% | 3.1% |
| Washington | 2.6% | 1.8% |
| Washington D. C. | 0.7% | 0.4% |
| West Virginia | 0.3% | 0.6% |
| Wisconsin | 2.0% | 2.1% |
| Wyoming | 0.1% | 0.2% |

**Table S1. Job Openings by State - Comparing Online Job Vacancies (OJV) Data to Bureau of Labor Statistics (BLS) Data.** Information leveraging the closest publicly available comparative data for job openings by state: *Job Openings by State (2023)*. (OJV data excludes "District of Columbia" which are included in the BLS dataset. Consequently, shares for BLS in the table are scaled accordingly, to ensure comparability.

| Occupation Category (SOC 2) | OJV Data | BLS Data |
|---|---|---|
| Architecture and Engineering Occupations | 2.3% | 1.6% |
| Arts, Design, Entertainment, Sports, and Media Occupations | 2.4% | 1.4% |
| Building and Grounds Cleaning and Maintenance Occupations | 2.1% | 3.0% |
| Business and Financial Operations Occupations | 7.6% | 6.5% |
| Community and Social Service Occupations | 1.6% | 1.7% |
| Computer and Mathematical Occupations | 8.4% | 3.3% |
| Construction and Extraction Occupations | 1.7% | 4.0% |
| Educational Instruction and Library Occupations | 2.9% | 7.2% |
| Farming, Fishing, and Forestry Occupations | 0.2% | 0.3% |
| Food Preparation and Serving Related Occupations | 4.9% | 8.6% |
| Healthcare Practitioners and Technical Occupations | 12.6% | 6.6% |
| Healthcare Support Occupations | 3.4% | 4.7% |
| Installation, Maintenance, and Repair Occupations | 3.9% | 3.8% |
| Legal Occupations | 0.6% | 0.8% |
| Life, Physical, and Social Science Occupations | 1.4% | 1.0% |
| Management Occupations | 9.9% | 6.8% |
| Office and Administrative Support Occupations | 9.3% | 12.1% |
| Personal Care and Service Occupations | 1.4% | 2.0% |
| Production Occupations | 3.4% | 5.4% |
| Protective Service Occupations | 1.2% | 2.4% |
| Sales and Related Occupations | 10.4% | 8.3% |
| Transportation and Material Moving Occupations | 6.8% | 8.6% |

**Table S2. Occupational Category Composition - Comparing Online Job Vacancies (OJV) Data to Bureau of Labor Statistics (BLS) Data**
Information leveraging the closest publicly available comparative data for job openings by occupational category: *Employment by occupational category (2023)*. (BLS data excludes “unclassified occupations” and “military occupations” which are included in the OJV dataset. Consequently, shares for OJV in the table are scaled accordingly, to ensure comparability).

## S2. Sample Construction

This section describes the procedures used for identifying AI roles and skills clusters.

### S2.A. Defining Skills

To classify complementary and substitutable skills, we consulted the literature (e.g., 9, 10, 11, 12, 13, 14). From this review, we form seven complementary skills clusters: “Analytical Thinking”, “Digital Literacy”, “Resilience”, “Technical Proficiency”, “Ethics”, “Working with Others”, and “Self-Efficiency”. For “substitutable” skills we extract four clusters: “Summary and Reporting”, “Language and Text Review”, “Customer Service”, and “Office and Financial Administration”. While the literature also proposes “Basic Data Skills”, “Information Management”, and “Prediction and Forecasting”, these are not included in our final “substitutable” skills clusters, as we believe these types of skills are likely to co-occur with AI requirements. To illustrate, “Prediction and Forecasting” may be required for data scientists using AI, and equally, AI researchers need coworkers to manage company information. The final skills clusters, alongside the corresponding literature references are presented in Tables S3 and S4.

| Skill | Literature Review (non-exhaustive) |
|---|---|
| Analytical thinking | (15, 16, 17, 18) |
| Digital literacy | (18, 19) |
| Resilience | (17, 18, 20) |
| Technical proficiency | (18, 21, 22) |
| Ethics | (23, 24, 25, 26) |
| Working with others | (15, 27, 18) |
| Self-efficiency | (27, 18) |

**Table S3. “Complementary” Skills Clusters.** Consulting the literature, we select seven skills clusters to capture AI-complementing skills in our analyses.

| Skill | Literature Review (non-exhaustive) |
|---|---|
| Summary and reporting | (28, 29) |
| Language and text review | (30, 31, 32) |
| Customer service | (33, 34, 35, 36) |
| Office and financial administration | (37, 38, 18) |

**Table S4. “Substitutable” Skills Clusters.** Consulting the literature, we select four skills clusters to capture AI-substitutable skills in our analyses.

We proceed by identifying specific skills within the Lightcast taxonomy. We embed all skills with more than 1000 related posts, as well as each of the category titles with "all-mpnet-base-v2" from the Sentence Transformer library. For each category, we select relevant skills from 10 skills with the highest cosine similarity. It should be highlighted that skills within a category have vastly different occurrence rates. Tables S5 and S6 list all of the skills found for the "complementary" and "substitutable" skills categories, as well as the number of postings mentioning each skill and our decision of whether or not to include a given skill in our analysis. If a job description includes at least one skill from a skill cluster, it is classified as requiring that given skill cluster. This threshold minimises sample loss (see Figure S1) and aligns with prior research (4, 39).

| Complementary Group | Skill | Count | Included |
|---|---|---|---|
| Analytical Thinking | Problem Solving | 1108795 | True |
| Analytical Thinking | Critical Thinking | 273688 | True |
| Analytical Thinking | Systems Thinking | 6764 | True |
| Analytical Thinking | Creative Thinking | 29065 | True |
| Analytical Thinking | Strategic Thinking | 56005 | True |
| Analytical Thinking | Analytical Skills | 266223 | True |
| Analytical Thinking | Independent Thinking | 12969 | True |
| Analytical Thinking | Intelligence Analysis | 4706 | True |
| Analytical Thinking | Creative Problem Solving | 37994 | True |
| Analytical Thinking | Analytical Thinking | 32338 | True |
| Digital Literacy | Digital Content | 8124 | True |
| Digital Literacy | Digital Arts | 1201 | False |
| Digital Literacy | Digital Capabilities | 2398 | True |
| Digital Literacy | Information Literacy | 1221 | True |
| Digital Literacy | Educational Technologies | 9478 | False |
| Digital Literacy | Literacy | 14192 | False |
| Digital Literacy | Digital Literacy | 4206 | True |
| Digital Literacy | Computer Literacy | 664990 | True |
| Digital Literacy | Health Literacy | 1321 | False |
| Digital Literacy | Digital Acumen | 1166 | True |
| Resilience | Tactfulness | 128776 | True |
| Resilience | Courage | 16828 | True |
| Resilience | Preparedness | 4426 | True |
| Resilience | Disaster Preparedness | 2564 | False |
| Resilience | Ingenuity | 9511 | True |
| Resilience | Tenacity | 15986 | True |
| Resilience | Healing | 1892 | False |
| Resilience | Resilience | 62862 | True |
| Resilience | Resourcefulness | 79478 | False |
| Resilience | Survivability | 1572 | True |
| Technical Proficiency | Study Skills | 2106 | False |
| Technical Proficiency | Technical Training | 42819 | True |
| Technical Proficiency | Analytical Skills | 266223 | False |
| Technical Proficiency | Mechanical Aptitude | 57640 | True |
| Technical Proficiency | Technical Subject Matter | 5436 | True |
| Technical Proficiency | Diagnostic Skills | 4111 | True |
| Technical Proficiency | Digital Acumen | 1166 | True |
| Technical Proficiency | Fine Motor Skills | 111080 | False |
| Technical Proficiency | Technical Acumen | 55124 | True |
| Technical Proficiency | Competency Assessment | 5333 | True |
| Ethics | Honesty | 120982 | True |
| Ethics | Confidentiality | 372250 | False |
| Ethics | Professional Responsibility | 3203 | True |
| Ethics | Business Ethics | 14730 | True |
| Ethics | Medical Ethics | 8246 | False |
| Ethics | Personal Integrity | 24973 | True |
| Ethics | Nursing Ethics | 1243 | False |
| Ethics | Ethical Standards And Conduct | 303842 | True |
| Ethics | Accountability | 330622 | True |
| Ethics | Compassion | 123864 | False |
| Working with Others | Interdisciplinary Collaboration | 2700 | True |
| Working with Others | Teamwork | 474835 | True |
| Working with Others | Delegation Skills | 13083 | True |
| Working with Others | Collaborative Communications | 1609 | True |
| Working with Others | Collaboration | 133059 | True |
| Working with Others | Team Building | 87719 | True |
| Working with Others | Cross-Functional Collaboration | 10986 | True |
| Working with Others | Team Processes | 4299 | True |
| Working with Others | Partner Development | 2623 | True |
| Working with Others | Support Colleagues | 1052 | True |
| Self-Efficiency | Self-Awareness | 14216 | True |
| Self-Efficiency | Self-Motivation | 413397 | True |
| Self-Efficiency | Self-Control | 6464 | True |
| Self-Efficiency | Self-Sufficiency | 4780 | True |
| Self-Efficiency | Self-Discipline | 81260 | True |
| Self-Efficiency | Business Efficiency | 1162 | True |
| Self-Efficiency | Conscientiousness | 17643 | True |
| Self-Efficiency | Resourcefulness | 79478 | True |
| Self-Efficiency | Operational Efficiency | 27450 | False |
| Self-Efficiency | Self-Regulation | 1157 | False |

**Table S5. “Complementary” Skills Clusters Matched with Lightcast Skills Taxonomy.** For each of the seven “complementary” skill clusters, we match skills from the Lightcast taxonomy, and record whether the given skill is included in the final set of skills used to identify “complementary” skill in the analysis. We further report the number of job postings mentioning each skill.

| Substitute Group | Skill | Count | Included |
|---|---|---|---|
| Summary and Reporting | Management Reporting | 32834 | True |
| Summary and Reporting | Report Creation | 2247 | True |
| Summary and Reporting | Report Development | 5148 | True |
| Summary and Reporting | Report Analysis | 1609 | True |
| Summary and Reporting | Statistical Reporting | 14162 | False |
| Summary and Reporting | Operational Reporting | 4375 | True |
| Summary and Reporting | Business Reporting | 6120 | True |
| Summary and Reporting | Report Building | 1312 | True |
| Summary and Reporting | Reporting and Analysis | 10849 | True |
| Summary and Reporting | Reports Analysis | 1345 | True |
| Language and Text Review | Language Arts | 5201 | True |
| Language and Text Review | Writing Systems | 27337 | True |
| Language and Text Review | Reviewing Applications | 2408 | True |
| Language and Text Review | Code Review | 40232 | False |
| Language and Text Review | Document Review | 8125 | True |
| Language and Text Review | English Language | 635617 | True |
| Language and Text Review | Foreign Language | 9874 | True |
| Language and Text Review | Language Development | 2150 | True |
| Language and Text Review | Multilingualism | 197313 | True |
| Language and Text Review | Written English | 22707 | True |
| Customer Service | Customer Service | 2622996 | True |
| Customer Service | Customer Service Desk | 3276 | True |
| Customer Service | Healthcare Customer Service | 1342 | True |
| Customer Service | Sales Support | 46062 | True |
| Customer Service | Client Services | 34742 | True |
| Customer Service | Support Services | 65689 | True |
| Customer Service | Technical Support | 96506 | True |
| Customer Service | Customer Support | 77539 | True |
| Customer Service | Customer Service Training | 3611 | True |
| Customer Service | Customer Service Management | 4955 | True |
| Office and Financial Administration | Financial Services | 125245 | False |
| Office and Financial Administration | Office Management | 63841 | True |
| Office and Financial Administration | Certified Management Accountant | 1103 | False |
| Office and Financial Administration | Bookkeeping | 71591 | True |
| Office and Financial Administration | Sales Administration | 3291 | True |
| Office and Financial Administration | Accounting Management | 3851 | True |
| Office and Financial Administration | Business Administration | 133334 | True |
| Office and Financial Administration | Office Administration | 17956 | True |
| Office and Financial Administration | Accounting | 472847 | False |
| Office and Financial Administration | Financial Management | 60884 | False |

**Table S6. “Substitutable” Skills Clusters Matched with Lightcast Skills Taxonomy.** For each of the four “substitutable” skill clusters, we match skills from the Lightcast taxonomy, and record whether the given skill is included in the final set of skills used to identify “substitutable” skill in the analysis. We further report the number of job postings mentioning each skill.

## S2.B. Defining AI-related Roles

The analysis relies on the distinction between AI-related and non-AI roles (i.e. positions working with AI technologies, and positions which do not. Following a skills-based approach (3, 4, 10), we identify AI-related roles by the presence of at least one AI skill requirement in the job posting. This method avoids misclassifying non-AI positions within AI-focused firms (e.g., a Human Resources Manager at an AI company) as AI roles.

We use the list of skills in Lightcast's "Artificial Intelligence and Machine Learning (AI/ML)" subcategory as our sample of AI skills, with the exception of the generic terms "Artificial Intelligence" and "Machine Learning", which may be associated with the company description. The complete set of skills in this subcategory, as well as the number of job postings mentioning each skill, and our decision on whether to include the skill in the analysis, are shown in Table S7. Figure S1 further displays the number of job postings requiring at least one "complementary", "substitutable", or "AI/ML" skill, as well as the total number of skills clusters that are mentioned.

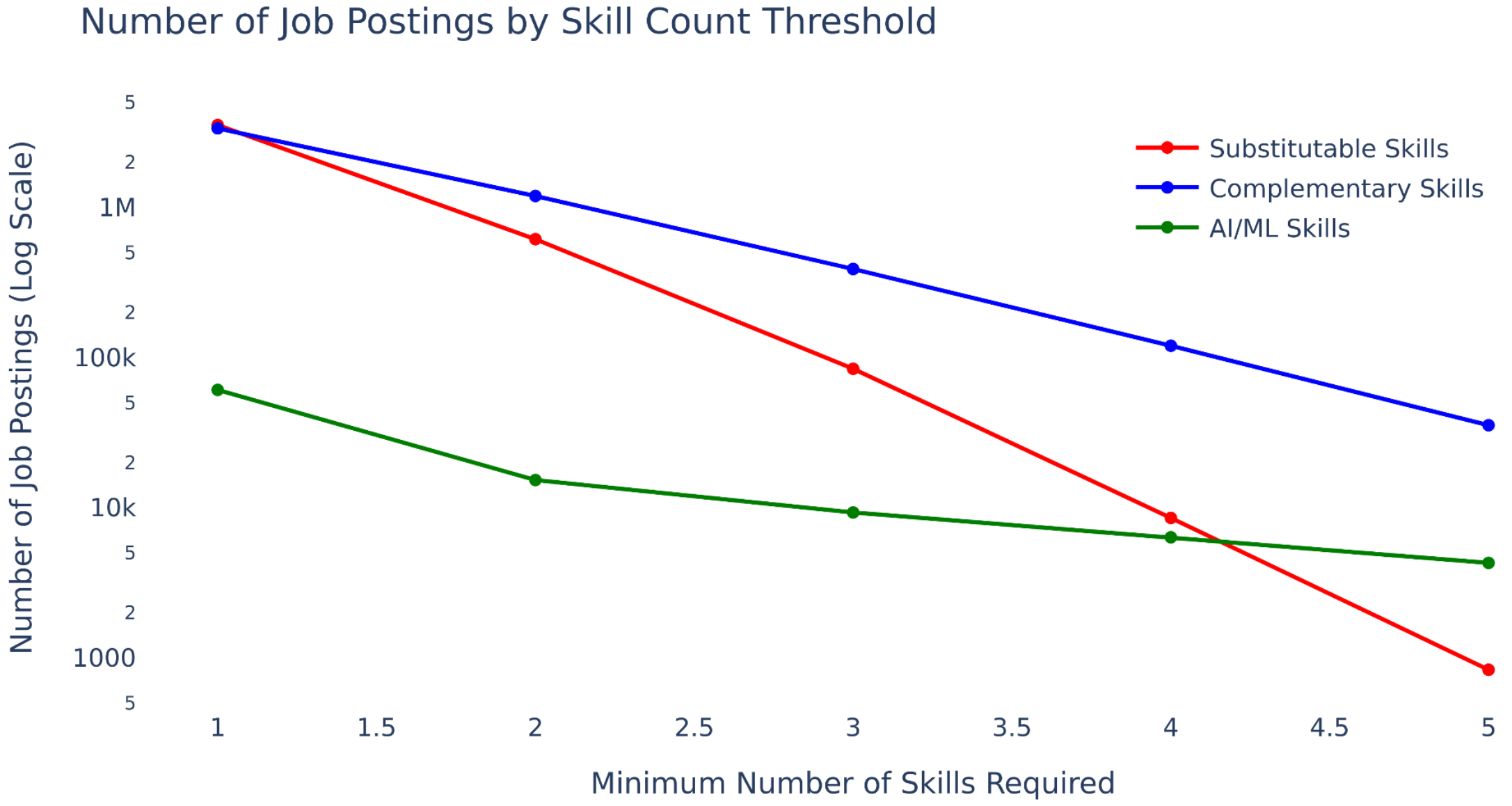

**Figure S1. The Count of Job Postings requiring "Complementary", "Substitutable" or "Artificial Intelligence/Machine Learning" Skills.** The graph further reveals how many individual skills clusters of each skill type are mentioned in job postings.

| AI Skill (w/ >200 count) | Count | Included |
|---|---|---|
| Machine Learning | 57774 | False |
| Artificial Intelligence | 46187 | False |
| Deep Learning | 8807 | True |
| TensorFlow | 6784 | True |
| Automation Systems | 6217 | True |
| Text Retrieval Systems | 5800 | True |
| Data Version Control (DVC) | 4705 | True |
| Machine Learning Algorithms | 4448 | True |
| PyTorch (Machine Learning Library) | 4350 | True |
| Artificial Neural Networks | 3382 | True |
| Scikit-Learn (Python Package) | 3249 | True |
| Generative Artificial Intelligence | 3015 | True |
| Chatbot | 2478 | True |
| Keras (Neural Network Library) | 2045 | True |
| MLOps (Machine Learning Operations) | 1571 | True |
| AWS SageMaker | 1505 | True |
| Random Forest Algorithm | 1273 | True |
| Intelligent Automation | 1192 | True |
| Reinforcement Learning | 1167 | True |
| Feature Engineering | 1088 | True |
| Machine Learning Methods | 1062 | True |
| Systems Automation | 1056 | True |
| Support Vector Machine | 1028 | True |
| Automated Data Cleaning | 959 | True |
| Knowledge-Based Systems | 923 | True |
| Large Language Modeling | 918 | True |
| OpenCV | 796 | True |
| Natural Language Understanding (NLU) | 768 | True |
| Azure Machine Learning | 767 | True |
| Boosting | 710 | True |
| Unsupervised Learning | 649 | True |
| AIOps (Artificial Intelligence For IT Operations) | 648 | True |
| Recommender Systems | 646 | True |
| Cognitive Computing | 616 | True |
| Torch (Machine Learning) | 610 | True |
| H2O.ai | 596 | True |
| Transformer (Machine Learning Model) | 550 | True |
| Object Tracking | 549 | True |
| Apache MXNet | 540 | True |
| Dynamic Routing | 533 | True |
| Applications Of Artificial Intelligence | 504 | True |
| Knowledge Engineering | 502 | True |
| AI Research | 501 | True |
| Recurrent Neural Network (RNN) | 488 | True |
| OpenAI Gym Environments | 476 | True |
| Kubeflow | 462 | True |
| Association Rule Learning | 455 | True |
| Federated Learning | 444 | True |
| Feature Extraction | 438 | True |
| Data Sovereignty | 426 | True |
| Edge Intelligence | 426 | True |
| Language Models | 425 | True |
| Xgboost | 421 | True |
| MLflow | 418 | True |
| Theano (Software) | 403 | True |
| Feature Selection | 395 | True |
| Intelligent Systems | 390 | True |
| General-Purpose Computing On Graphics Processing Units | 363 | True |
| Knowledge-Based Configuration | 349 | True |
| Machine Learning Model Training | 345 | True |
| Expert Systems | 343 | True |
| Generative Adversarial Networks | 325 | True |
| Text to Speech (TTS) | 324 | True |
| Supervised Learning | 322 | True |
| ChatGPT | 315 | True |
| Gradient Boosting | 308 | True |
| Artificial Intelligence Development | 307 | True |
| Support Vector Machines (SVM) | 306 | True |
| Automated Machine Learning | 295 | True |
| Convolutional Neural Networks (CNN) | 292 | True |
| Distributed Machine Learning | 288 | True |
| Prompt Engineering | 282 | True |
| Dask (Software) | 277 | True |
| Knowledge Representation | 273 | True |
| Multi-Agent Systems | 272 | True |
| Artificial Intelligence Markup Language (AIML) | 271 | True |
| Apache Mahout | 269 | True |
| Concept Drift Detection | 267 | True |
| Intelligent Virtual Assistant | 263 | True |
| Bagging Techniques | 259 | True |
| Long Short-Term Memory (LSTM) | 252 | True |
| Ensemble Methods | 251 | True |
| AI/ML Inference | 242 | True |
| Transfer Learning | 238 | True |
| Deep Learning Methods | 237 | True |
| Voice Assistant Technology | 232 | True |
| ModelOps | 206 | True |

**Table S7. Skills Used to Identify AI-related Roles from Job Descriptions.**
The set of skills is drawn from Lightcasts' "Artificial Intelligence and Machine Learning (AI/ML)" subcategory. "Machine Learning" and "Artificial Intelligence" are excluded from the set of skills used to construct the final sample of AI-related roles for the analysis. We further report the number of job descriptions mentioning each skill with at least 200 instances.

## S3. Applied Methods

This section outlines the statistical models used to estimate internal and external effects, including logistic and linear regressions, as well as the control variables employed.

### S3.A. Internal Effects Analysis

Regression as a method of analysis has an established precedent of being used to understand the relationships between artificial intelligence and patterns demand patterns in the labour market (e.g. 1, 40). In this paper, we analyse the internal effects by employing a logistic regression to understand the associations between AI roles and demand for complementary and substitutable skills. The set of control variables included in this regression analysis is provided in Table S8.

The regression formula is captured as follows:

$$\log(\frac{p}{1-p}) = \beta_0 + \beta_1 X_{1,i} + \delta \mathbf{X}_i' + \epsilon_i$$

Where:

- $p = Pr(Y_i = 1)$ is the probability of a job posting advertising a given "complementary" skill $Y_i$
- $\beta_0$ is the intercept
- $\beta_1$ captures the effect of a job posting being AI-related
- $X_{1,i} = 1$ indicates that job posting $i$ is AI-related
- $X_i'$ is a vector of control variables, as listed in Table S8

The complete set of results is presented in Table S9. We also provide results for the demand for "substitutable" skills in Table S10.

After this main analysis of the internal effects, we proceed to explore whether the demand for skills is reflected in the salaries paid in certain roles. We use a linear regression to understand how the presence of "complementary" or "substitutable" skills affect the (log-transformed) salary in AI-related roles. We again include the control variables listed in Table S8.

The regression formula for the wage analysis is written as:

$$\log(Y_i) = \beta_0 + \beta_1 X_{1,i} + \delta \mathbf{X}_i' + \epsilon_i$$

Where:

- $log(Y_i)$ is the salary (log-transformed) for a given role $i$
- $\beta_0$ is the intercept

- $\beta_1$ captures the effect of job-posting $i$ requiring a given complementary or substitutable skill
- $X_{1,i} = 1$ if job-posting $i$ requires a given complementary or substitutable skill
- $X_i$' is a vector of control variables, as listed in Table S8.

Tables S11 and S12 report the regression results for this wage analysis.

### S3.B. External Effects

To investigate the effect of AI prevalence on the demand of "complementary" and "substitutable" skill in non-AI roles, we employ a fractional logistic model. We aggregate the job postings data by year and conduct analyses at the company, industry, and regional level. Thus, we are able to measure AI prevalence as the share of job postings within a given entity, which are classified as AI-related roles. The main external effects analysis is conducted on the sample of companies, industries, or regions with at least one AI-related job posting. A set of control variables is listed in Table S8.

The regression equation is expressed as follows:

$$\log(\frac{p}{1-p}) = \beta_0 + \beta_1 X_{1,i} + \delta \mathbf{X}'_i + \epsilon_i$$

Where:

- $p = E(y_{i,k}|X_{i,k})$ is the predicted share of Non-AI related job postings requiring "complementary" or "substitutable" skills in year $i$ and company, industry, or region $k$
- $X_{1,i,k}$ captures AI penetration (share of job postings demanding AI skills) by year $i$ and company, industry, or region level $k$
- $\beta_0$ is the intercept
- $\beta_1$ is the coefficient on AI prevalence. The sign of the coefficient indicates the direction of the effect of AI prevalence on demand for "complementary" or "substitutable" skills in non-AI roles ($Y_{i,k}$)
- $\epsilon_{i,k}$ is the error term

At the company-level, we observe an inflated number of companies with zero AI roles posted for all three geographies. This zero-inflation further allows us to extend our main analysis to the extensive margin. Hence, we can capture how the demand for "complementary" or "substitutable" skills is affected by whether or not the company has any AI-related postings. In this case, our regression equation presented above remains unchanged, except that our main explanatory variable is now expressed as follows:

$$X_{1,i,k} = \begin{cases} 1, & \text{company } k \text{ has} \geq 1 \text{ AI postings} \\ 0, & \text{company } k \text{ has } 0 \text{ AI postings} \end{cases}$$

The complete set of results, at the company, industry, and region levels for the United States, United Kingdom, Australia and New Zealand are presented in Tables S13. The extensive margin results at the company level are further displayed in Table S14.

| Control Variable | Description/Operationalisation | Justification |
|---|---|---|
| Minimum years of experience *(internal effects)* | Minimum years of experience required. Discrete variable as specified by job posting. | Salary differs by experience level and different experience levels are associated with different skills demands (41, 42, 43) |
| Minimum education level *(internal effects)* | Minimum education level, as specified by job posting. (None listed, High School or GED, Associate's Degree, Bachelor's Degree, Master's Degree, PhD or professional degree) (ordinal variable) | Salaries differ by education level and skills demand is related to education levels (44, 4, 45, 46) |
| Year *(internal and external effects)* | Year (control variables for 2019-2023 included, reference year is the omitted category 2018) | Labour/skills demand and salary fluctuate and differ year to year (44, 45) |
| Region *(internal effects)* | Controlling for regional fixed effects. | |
| Industry *(internal and external effects at company-level)* | Controlling for industry-specific fixed effects. | |
| Total postings *(external effects)* | Total number of job postings per year per observation category (company, industry or region) | Captures overall labour market activity, ensuring that the observed effects of AI role postings are not confounded by broader variations in labour demand (e.g., market size fluctuations). |

**Table S8. List of Control Variables Included in the Internal and External Effects Regression Analyses.** The choice of control variables is based on a literature review, as outlined in the third column.

## S4. Results

### S4.A. Descriptive Evidence

This section presents descriptive evidence of the demand for “complementary” and “substitutable” skills between 2018 and 2024. In the main manuscript, Figure 1 plots the odds-ratios for each of the seven “complementary” skills clusters over this time period. The graph reveals that five out of seven clusters are consistently more commonly demanded in AI-related roles, as opposed to non-AI roles. Here, Figure S2 replicates the odds-ratio calculations for the five “substitutable” skills clusters.

Motivated by recent findings (e.g., 49, 18) suggesting a recent major step change in AI adoption, specifically the use of (generative) AI to substitute and complement knowledge work, we further investigate the *change* in demand for complementary and substitutable skills clusters across AI role categories over time. The left panel of Figure S3 plots the share of AI-related and non-AI roles which require at least one of the skills clusters associated with the corresponding “complementary” or “substitutable” skills categories between 2018 and 2024. We further plot the median salaries associated with different skills categories and AI versus non-AI related roles in the right panel of the figure.

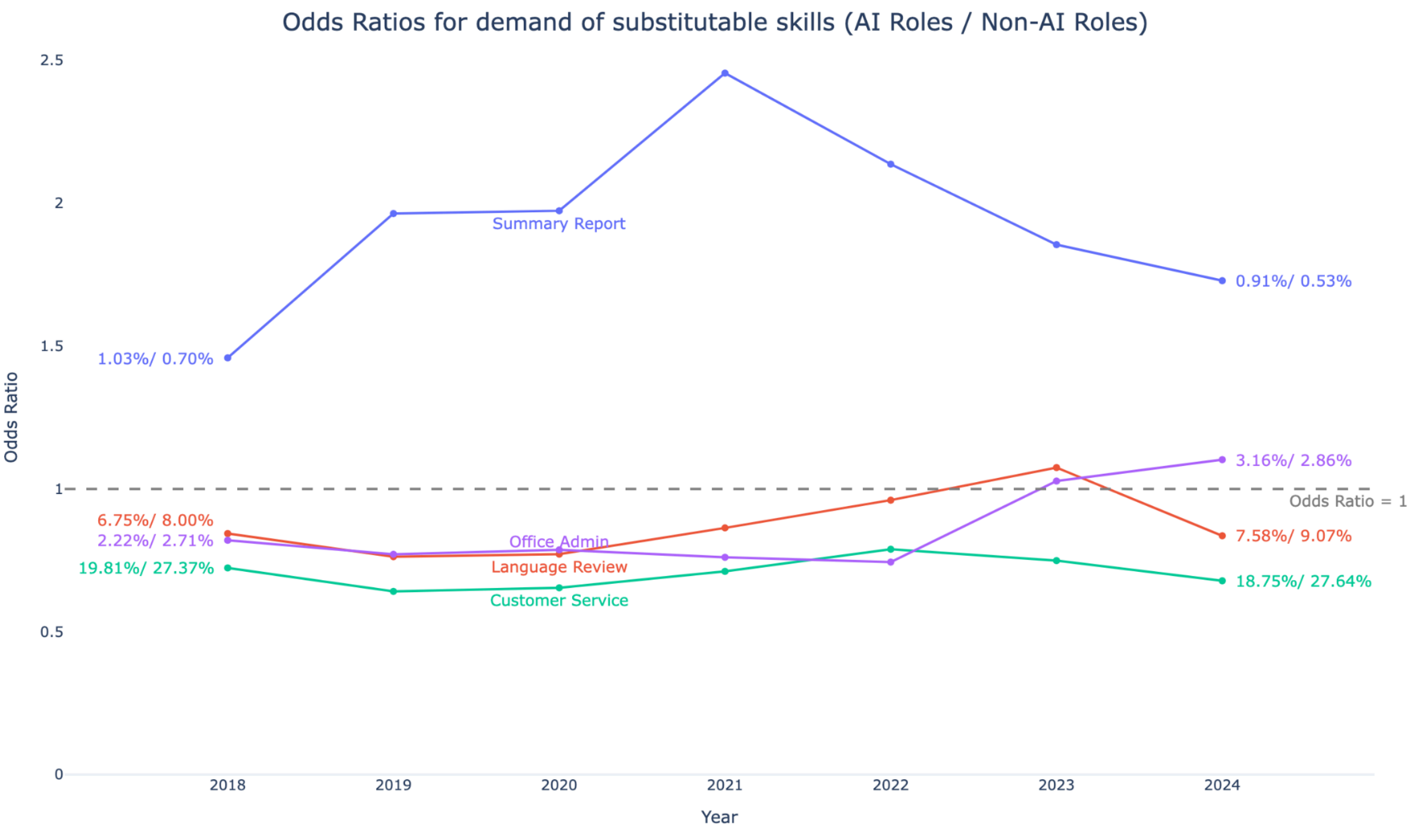

**Figure S2. Calculated odds-ratios relating the demand for “substitutable” skills in AI-related vs non-AI roles.** “Customer Service” skills are consistently less common in AI roles. “Language Review” and “Office Admin” skills are also less common in AI roles until the later years of the sample (2023-2024), where they become more commonly demanded in AI-related roles. Meanwhile, “summary report” skills are consistently higher demanded in AI-related roles than non-AI roles.

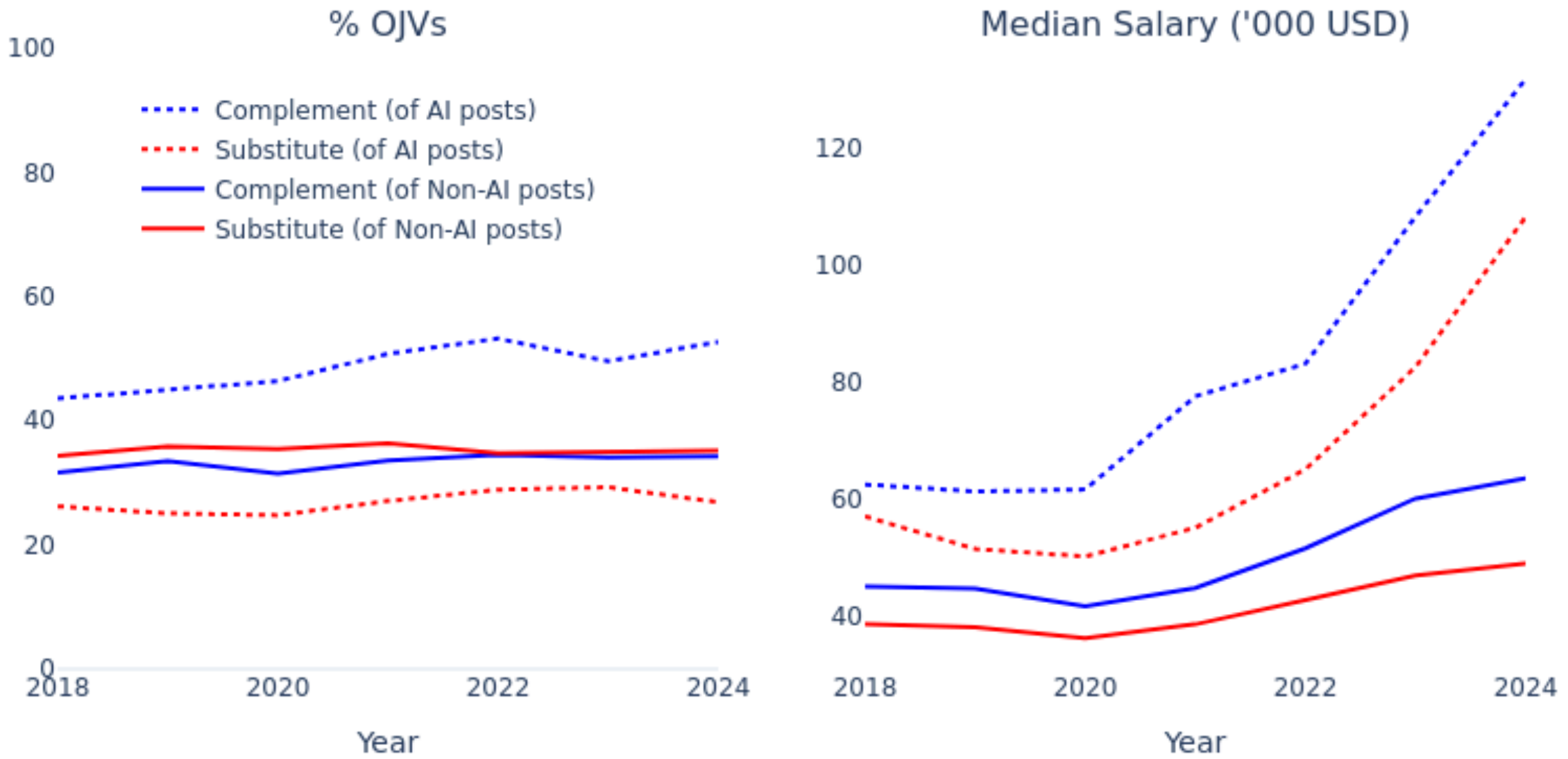

**Figure S3. Skills demand and median salaries across job postings between 2018 and 2024.** The left panel plots the demand for any "complementary" or "substitutable" skill in AI-related and non-AI roles. Demand for "complementary" skills is consistently higher than 40%, and growing between 2018 and 2024, while demand for "substitutable" skills remains between 20 and 30% amongst AI-related roles. In non-AI roles, "substitutable" skills are marginally more demanded than "complementary" skills, and the demand converges after 2022. The right panel plots the median salaries in job roles and reveals that while salaries are generally increasing, they are consistently highest in AI-related roles, specifically those demanding "complementary" skills. Median salaries are lowest and experience the least growth in non-AI roles, although those roles requiring "complementary" skills are again associated with higher salaries here.

### S4.B. Internal Effects Results

In this section we provide results from the internal effects analysis, beyond the findings presented in Figure 2 in the main manuscript. First, we explore how working with AI technologies affects the demand for "substitutable" skills within roles. We then proceed to modelling monetary valuation of skills by running a linear regression to capture how skills requirements affect the salary paid in AI-related roles. The resulting coefficients are presented visually in Figures S4 and S5.

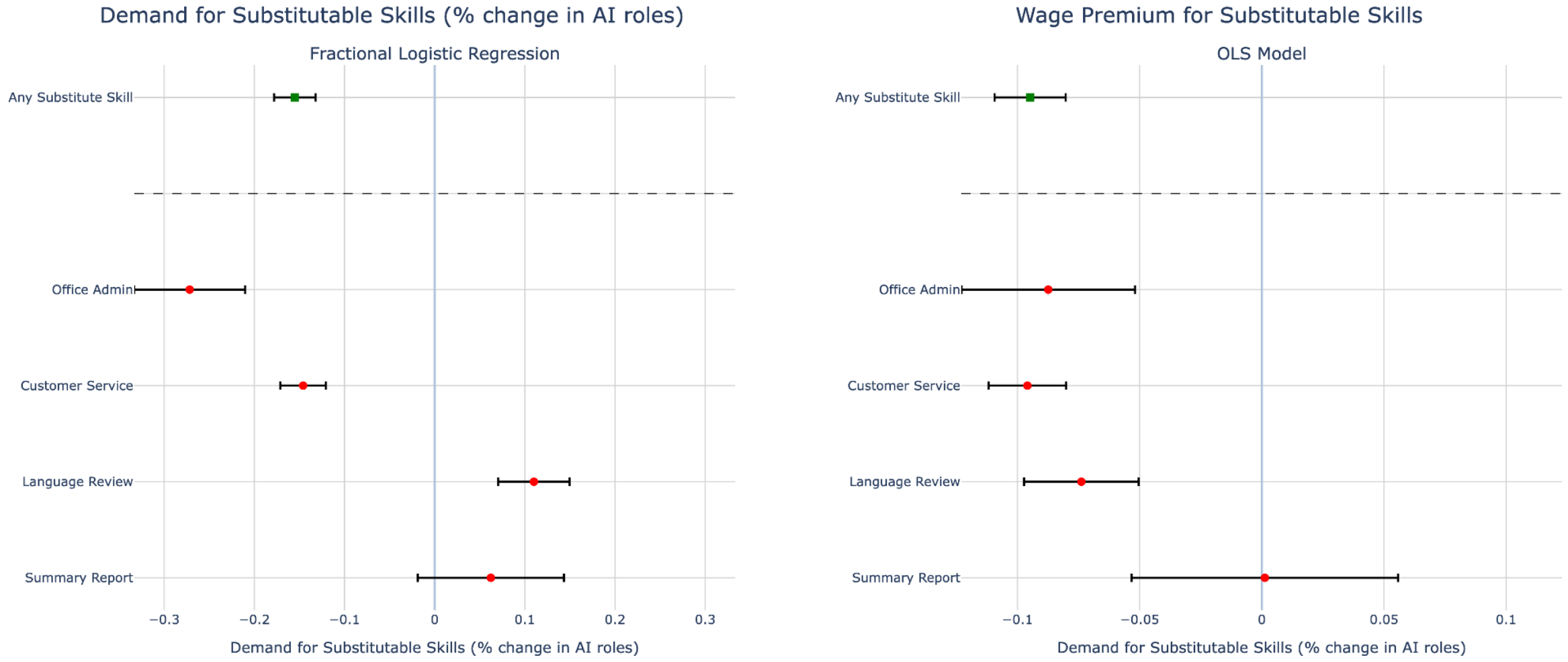

**Figure S4. Visual Representation of Internal Effects Regression Results for "Substitutable" Skills.** The left panel presents the effects of working with AI on the demand for "substitutable" skills. The right panel presents coefficients from linear regressions (OLS method) relating the presence of "substitutable" skills in AI-related roles to (log) salaries paid.

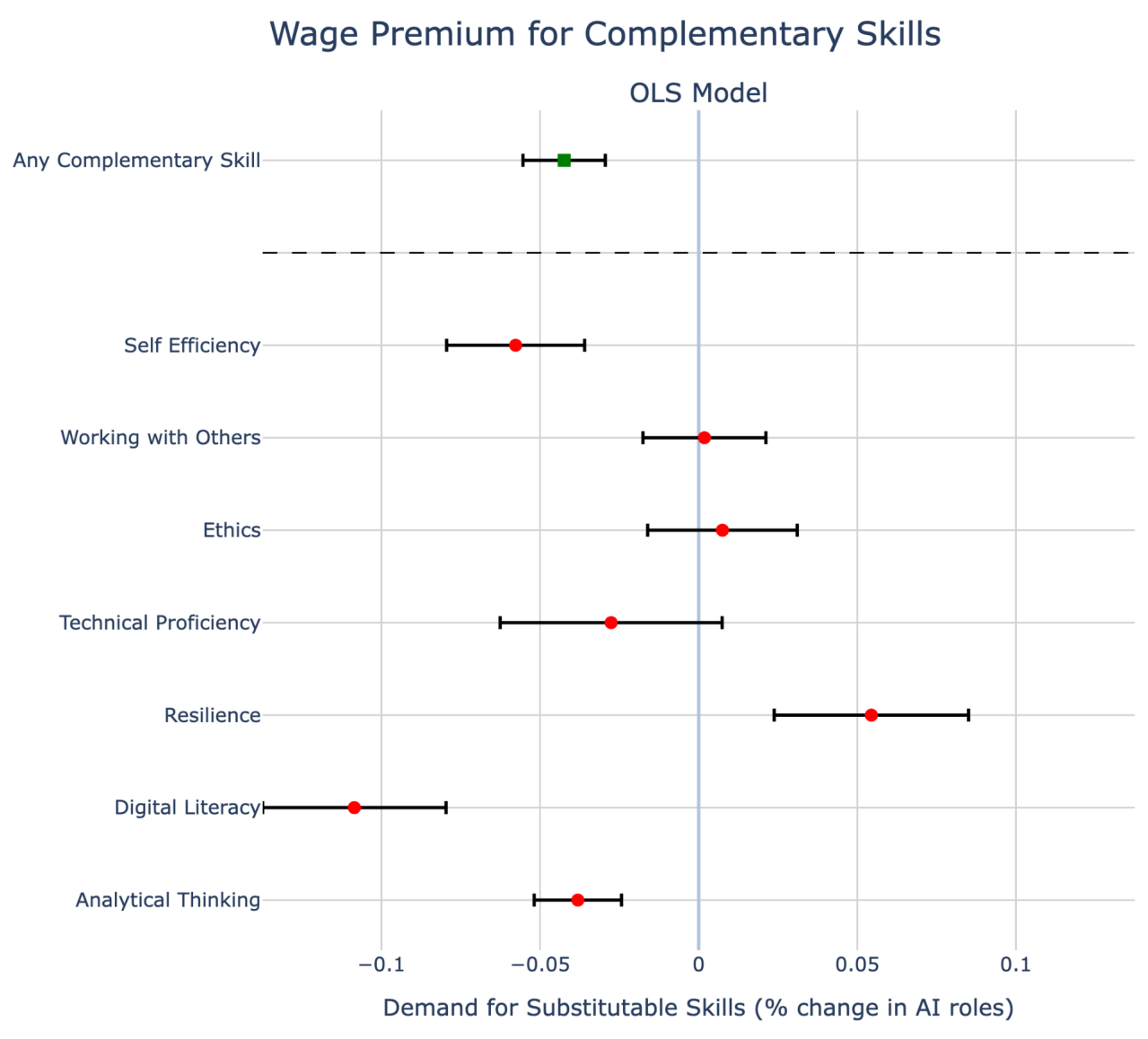

**Figure S5. Visual Representation of Internal Effects Wage Regressions Results for "Complementary" Skills.** Coefficients from linear regressions (OLS method) relating the presence of "complementary" skills in AI-related roles to (log) salaries paid.

We further provide the complete regression results for all internal effects analyses on "complementary" and "substitutable" skills. For "complementary" skills, as presented in Figure 2 of the main manuscript, the demand analysis is conducted for all occupations, as well as selected ONET occupation categories: management, business and financial operations, sales and related, and office and administrative support occupations. The complete results are displayed in Table S9.

Regression results for the wage premium analysis, and results for "substitutable" skills are further reported in Tables S10, S11, and S12.

As reported in Table S9 and Figure 2 in the main manuscript, the selection and extent of skills demanded further varies when selecting on occupations, particularly those where the potential for human-specific value added is expected to be rather high. We further explore this concept in Figure S6. Similar to Figure S2, this plots the odds-ratio for different “complementary” skills, although here we track how this relative demand changes with seniority of the job roles. There can be read a general increase in demand for “complementary” skills with years of experience. In positions demanding less experience, the majority of “complementary” skills clusters are more frequently demanded in AI-related than non-AI roles, though this difference drastically diminishes in more senior positions. The figure further compares salaries paid to roles with or without the given “complementary” skills, though no clear pattern can be identified here.

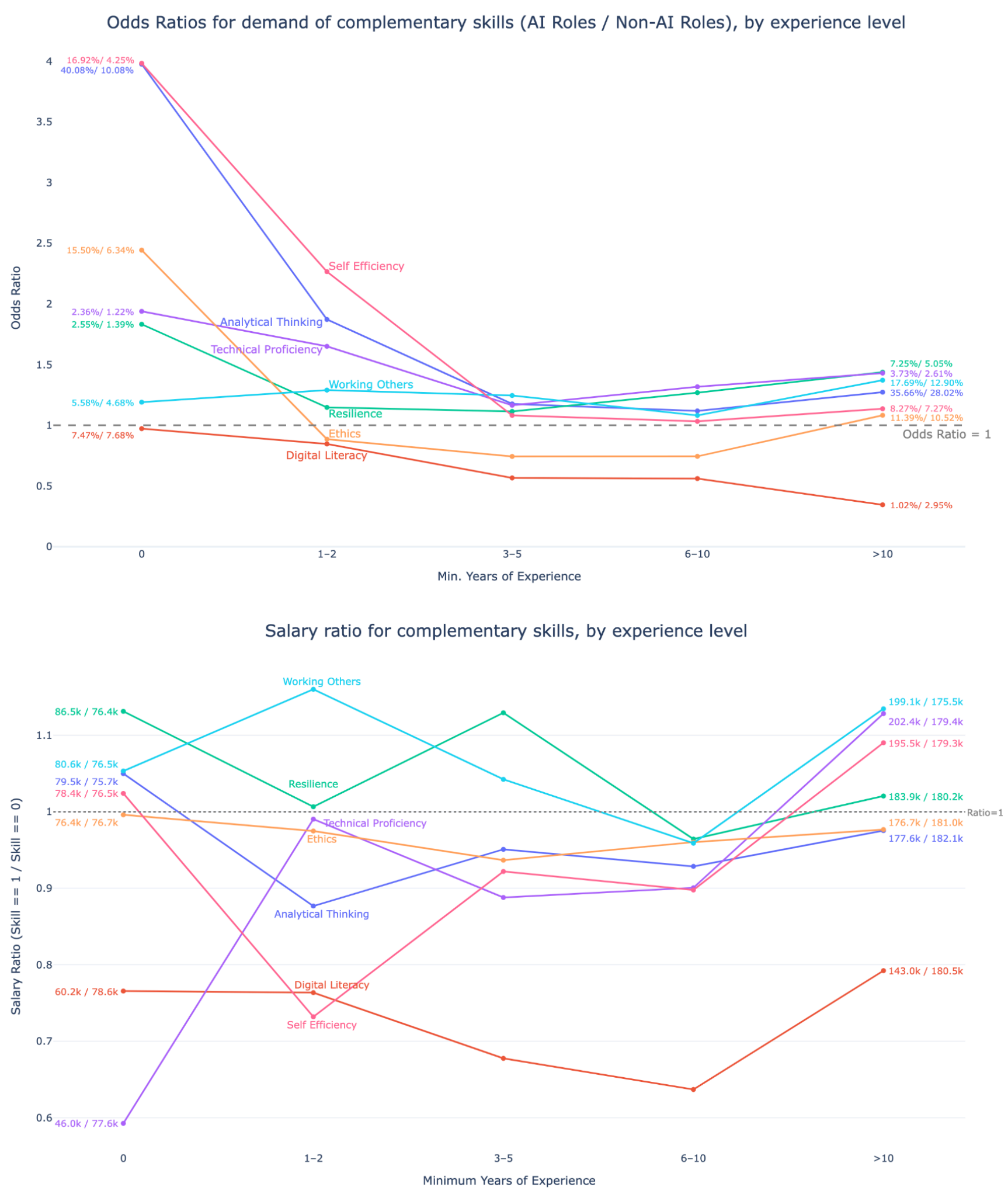

**Figure S6. Tracking changes in demand for “complementary” skills with seniority levels of roles.** The upper panel plots the odds ratio of demand for skills in AI-related versus non-AI roles across different experience levels. Overall, the difference in demand for “complementary” skills between roles working directly with or without AI technologies diminishes with seniority. The lower panel displays the ratio in salaries paid for different levels of experience, contrasting roles requiring a given “complementary” skill with ones that do not.

| Dependent variables: | Analytical Thinking | Digital Literacy | Resilience | Technical Proficiency | Ethics | Working with Others | Self Efficiency | Any Complementary Skill |
|---|---|---|---|---|---|---|---|---|
| **All Occupations** | | | | | | | | |
| AI | 0.17*** | -0.00 | 0.11*** | 0.24*** | -0.02 | 0.11*** | 0.22*** | 0.18*** |
| | (0.01) | (0.02) | (0.03) | (0.03) | (0.02) | (0.02) | (0.02) | (0.01) |
| Years of Experience (min) | 0.03*** | -0.04*** | 0.05*** | 0.03*** | 0.04*** | 0.05*** | -0.00*** | 0.03*** |
| | (0.00) | (0.00) | (0.00) | (0.00) | (0.00) | (0.00) | (0.00) | (0.00) |
| Intercept | -1.82*** | -2.05*** | -3.99*** | -3.39*** | -2.82*** | -2.67*** | -3.07*** | -0.66*** |
| | (0.02) | (0.02) | (0.04) | (0.04) | (0.02) | (0.02) | (0.03) | (0.01) |
| Fixed Effects: | | | | | | | | |
| Minimum education level | ✓ | ✓ | ✓ | ✓ | ✓ | ✓ | ✓ | ✓ |
| Year | ✓ | ✓ | ✓ | ✓ | ✓ | ✓ | ✓ | ✓ |
| Region | ✓ | ✓ | ✓ | ✓ | ✓ | ✓ | ✓ | ✓ |
| Industry | ✓ | ✓ | ✓ | ✓ | ✓ | ✓ | ✓ | ✓ |
| Occupation | ✓ | ✓ | ✓ | ✓ | ✓ | ✓ | ✓ | ✓ |
| N | 4520774 | 4520774 | 4520774 | 4520774 | 4520774 | 4520774 | 4520774 | 4520774 |
| Pseudo $R^2$ | 0.063 | 0.055 | 0.034 | 0.079 | 0.043 | 0.031 | 0.029 | 0.051 |
| **Management (ONET 11)** | | | | | | | | |
| AI | 0.32*** | 0.15*** | 0.12* | 0.42*** | -0.10** | 0.20*** | 0.28*** | 0.28*** |
| | (0.03) | (0.06) | (0.07) | (0.09) | (0.05) | (0.04) | (0.05) | (0.03) |
| Years of Experience (min) | 0.04*** | -0.08*** | 0.03*** | 0.02*** | 0.04*** | 0.05*** | -0.01*** | 0.03*** |
| | (0.00) | (0.00) | (0.00) | (0.00) | (0.00) | (0.00) | (0.00) | (0.00) |
| Intercept | -1.79*** | -1.48*** | -3.89*** | -4.25*** | -2.07*** | -2.41*** | -3.13*** | -0.51*** |
| | (0.04) | (0.05) | (0.08) | (0.12) | (0.05) | (0.05) | (0.06) | (0.03) |
| Fixed Effects: | | | | | | | | |
| Minimum education level | ✓ | ✓ | ✓ | ✓ | ✓ | ✓ | ✓ | ✓ |
| Year | ✓ | ✓ | ✓ | ✓ | ✓ | ✓ | ✓ | ✓ |
| Region | ✓ | ✓ | ✓ | ✓ | ✓ | ✓ | ✓ | ✓ |
| Industry | ✓ | ✓ | ✓ | ✓ | ✓ | ✓ | ✓ | ✓ |
| Occupation | ✗ | ✗ | ✗ | ✗ | ✗ | ✗ | ✗ | ✗ |
| N | 616251 | 616251 | 616251 | 616251 | 616251 | 616251 | 616251 | 616251 |
| Pseudo $R^2$ | 0.026 | 0.040 | 0.015 | 0.017 | 0.011 | 0.015 | 0.012 | 0.018 |
| **Business & Financial Operations (ONET 13)** | | | | | | | | |
| AI | 0.15*** | -0.19*** | -0.19** | 0.32*** | -0.05 | 0.26*** | 0.15*** | 0.14*** |
| | (0.03) | (0.07) | (0.09) | (0.10) | (0.05) | (0.05) | (0.05) | (0.03) |
| Years of Experience (min) | 0.01*** | -0.08*** | 0.04*** | 0.05*** | 0.05*** | 0.05*** | -0.03*** | 0.01*** |
| | (0.00) | (0.00) | (0.00) | (0.00) | (0.00) | (0.00) | (0.00) | (0.00) |
| Intercept | -1.10*** | -1.41*** | -3.35*** | -3.23*** | -2.75*** | -2.36*** | -2.15*** | -0.06 |
| | (0.05) | (0.07) | (0.12) | (0.14) | (0.08) | (0.07) | (0.07) | (0.04) |
| Fixed Effects: | | | | | | | | |
| Minimum education level | ✓ | ✓ | ✓ | ✓ | ✓ | ✓ | ✓ | ✓ |
| Year | ✓ | ✓ | ✓ | ✓ | ✓ | ✓ | ✓ | ✓ |
| Region | ✓ | ✓ | ✓ | ✓ | ✓ | ✓ | ✓ | ✓ |
| Industry | ✓ | ✓ | ✓ | ✓ | ✓ | ✓ | ✓ | ✓ |
| Occupation | ✗ | ✗ | ✗ | ✗ | ✗ | ✗ | ✗ | ✗ |
| N | 478266 | 478266 | 478266 | 478266 | 478266 | 478266 | 478266 | 478266 |
| Pseudo $R^2$ | 0.020 | 0.029 | 0.017 | 0.015 | 0.017 | 0.016 | 0.008 | 0.018 |
| **Sales & Related (ONET 41)** | | | | | | | | |
| AI | 0.83*** | -0.11 | -0.33** | 0.22 | 0.34*** | -0.19** | 0.99*** | 0.76*** |
| | (0.06) | (0.10) | (0.16) | (0.15) | (0.08) | (0.10) | (0.06) | (0.06) |
| Years of Experience (min) | 0.09*** | -0.06*** | -0.01** | 0.06*** | 0.02*** | 0.07*** | -0.01*** | 0.05*** |
| | (0.00) | (0.00) | (0.00) | (0.00) | (0.00) | (0.00) | (0.00) | (0.00) |
| Intercept | -2.08*** | -1.85*** | -3.74*** | -4.61*** | -1.39*** | -2.73*** | -3.24*** | -0.34*** |
| | (0.05) | (0.06) | (0.10) | (0.17) | (0.06) | (0.07) | (0.08) | (0.04) |
| Fixed Effects: | | | | | | | | |
| Minimum education level | ✓ | ✓ | ✓ | ✓ | ✓ | ✓ | ✓ | ✓ |
| Year | ✓ | ✓ | ✓ | ✓ | ✓ | ✓ | ✓ | ✓ |
| Region | ✓ | ✓ | ✓ | ✓ | ✓ | ✓ | ✓ | ✓ |
| Industry | ✓ | ✓ | ✓ | ✓ | ✓ | ✓ | ✓ | ✓ |
| Occupation | ✗ | ✗ | ✗ | ✗ | ✗ | ✗ | ✗ | ✗ |
| N | 388744 | 388744 | 388744 | 388744 | 388744 | 388744 | 388744 | 388744 |
| Pseudo $R^2$ | 0.049 | 0.031 | 0.022 | 0.044 | 0.040 | 0.019 | 0.029 | 0.021 |
| **Office & Administration Support (ONET 43)** | | | | | | | | |
| AI | 1.06*** | -0.18** | 0.07 | 0.17 | 0.01 | 0.28*** | 1.63*** | 0.88*** |
| | (0.05) | (0.08) | (0.12) | (0.19) | (0.10) | (0.09) | (0.06) | (0.06) |
| Years of Experience (min) | 0.08*** | -0.05*** | 0.08*** | 0.05*** | 0.07*** | 0.10*** | 0.04*** | 0.07*** |
| | (0.00) | (0.00) | (0.00) | (0.01) | (0.00) | (0.00) | (0.00) | (0.00) |
| Intercept | -1.78*** | -1.70*** | -3.25*** | -5.53*** | -2.80*** | -2.26*** | -3.21*** | -0.53*** |
| | (0.05) | (0.05) | (0.10) | (0.20) | (0.08) | (0.07) | (0.08) | (0.04) |
| Fixed Effects: | | | | | | | | |
| Minimum education level | ✓ | ✓ | ✓ | ✓ | ✓ | ✓ | ✓ | ✓ |
| Year | ✓ | ✓ | ✓ | ✓ | ✓ | ✓ | ✓ | ✓ |
| Region | ✓ | ✓ | ✓ | ✓ | ✓ | ✓ | ✓ | ✓ |
| Industry | ✓ | ✓ | ✓ | ✓ | ✓ | ✓ | ✓ | ✓ |
| Occupation | ✗ | ✗ | ✗ | ✗ | ✗ | ✗ | ✗ | ✗ |
| N | 409570 | 409570 | 409570 | 409570 | 409570 | 409570 | 409570 | 409570 |
| Pseudo $R^2$ | 0.034 | 0.021 | 0.036 | 0.016 | 0.025 | 0.031 | 0.015 | 0.023 |

Standard errors in parentheses. * p<0.1, ** p<0.05, ***p<0.01

**Table S9. Regression Results for “Complementary” Skills Demand.** Results from logistic regressions capturing the effect of working with AI technologies (AI-related roles) on demand for “complementary” skills. The analysis is performed on the complete sample of job postings, as well as subsamples of management, business and finance, sales, and office and administrative support occupations.

| Dependent variables: | Summary Report | Language Review | Customer Service | Office Admin | Any Substitutable Skill |
|---|---|---|---|---|---|
| | | | All Occupations | | |
| AI | 0.06 | 0.11*** | -0.15*** | -0.27*** | -0.16*** |
| | (0.04) | (0.02) | (0.01) | (0.03) | (0.01) |
| Years of Experience (min) | 0.03*** | -0.05*** | -0.04*** | 0.04*** | -0.04*** |
| | (0.00) | (0.00) | (0.00) | (0.00) | (0.00) |
| Intercept | -5.59*** | -2.80*** | -0.95*** | -4.87*** | -0.82*** |
| | (0.07) | (0.02) | (0.02) | (0.04) | (0.01) |
| Fixed Effects: | | | | | |
| Minimum education level | ✓ | ✓ | ✓ | ✓ | ✓ |
| Year | ✓ | ✓ | ✓ | ✓ | ✓ |
| Region | ✓ | ✓ | ✓ | ✓ | ✓ |
| Industry | ✓ | ✓ | ✓ | ✓ | ✓ |
| Occupation | ✓ | ✓ | ✓ | ✓ | ✓ |
| N | 4520774 | 4520774 | 4520774 | 4520774 | 4520774 |
| Pseudo $R^2$ | 0.086 | 0.050 | 0.123 | 0.138 | 0.105 |

Standard errors in parentheses. * $p<0.1$, ** $p<0.05$, *** $p<0.01$

**Table S10. Regression Results for “Substitutable” Skills Demand.** Results from logistic regressions capturing the effect of working with AI technologies (AI-related roles) on demand for “substitutable” skills.

Dependent variable: Salary (log)

| Skills premium for: | Analytical Thinking | Digital Literacy | Resilience | Technical Proficiency | Ethics | Working with Others | Self Efficiency | Any Complementary Skill |
|---|---|---|---|---|---|---|---|---|
| | | | | All Occupations | | | | |
| Skill's premium | -0.04*** | -0.11*** | 0.05*** | -0.03 | 0.01 | 0.00 | -0.06*** | -0.04*** |
| | (0.01) | (0.01) | (0.02) | (0.02) | (0.01) | (0.01) | (0.01) | (0.01) |
| Years of Experience (min) | 0.10*** | 0.10*** | 0.10*** | 0.10*** | 0.10*** | 0.10*** | 0.10*** | 0.10*** |
| | (0.00) | (0.00) | (0.00) | (0.00) | (0.00) | (0.00) | (0.00) | (0.00) |
| Intercept | 10.75*** | 10.76*** | 10.74*** | 10.75*** | 10.74*** | 10.74*** | 10.74*** | 10.76*** |
| | (0.05) | (0.05) | (0.05) | (0.05) | (0.05) | (0.05) | (0.05) | (0.05) |
| Fixed Effects: | | | | | | | | |
| Minimum education level | ✓ | ✓ | ✓ | ✓ | ✓ | ✓ | ✓ | ✓ |
| Year | ✓ | ✓ | ✓ | ✓ | ✓ | ✓ | ✓ | ✓ |
| Region | ✓ | ✓ | ✓ | ✓ | ✓ | ✓ | ✓ | ✓ |
| Industry | ✓ | ✓ | ✓ | ✓ | ✓ | ✓ | ✓ | ✓ |
| Occupation | ✓ | ✓ | ✓ | ✓ | ✓ | ✓ | ✓ | ✓ |
| R-squared | 0.67 | 0.67 | 0.67 | 0.67 | 0.67 | 0.67 | 0.67 | 0.67 |
| R-squared Adj. | 0.67 | 0.67 | 0.67 | 0.67 | 0.67 | 0.67 | 0.67 | 0.67 |
| N | 9990 | 9990 | 9990 | 9990 | 9990 | 9990 | 9990 | 9990 |

Standard errors in parentheses. * $p<0.1$, ** $p<0.05$, *** $p<0.01$

**Table S11. Regression Results for Wage Effects for “Complementary” Skills.** Results from linear regressions relating the requirement for “complementary” skills to wages offered within AI-related roles.

Dependent variable: Salary (log)

| Skills premium for: | Summary Report | Language Review | Customer Service | Office Admin | Any Substitutable Skill |
|---|---|---|---|---|---|
| | | | All Occupations | | |
| Skill's premium | 0.00 | -0.07*** | -0.10*** | -0.09*** | -0.09*** |
| | (0.03) | (0.01) | (0.01) | (0.02) | (0.01) |
| Years of Experience (min) | 0.10*** | 0.10*** | 0.10*** | 0.10*** | 0.10*** |
| | (0.00) | (0.00) | (0.00) | (0.00) | (0.00) |
| Intercept | 10.74*** | 10.75*** | 10.77*** | 10.74*** | 10.78*** |
| | (0.05) | (0.05) | (0.05) | (0.05) | (0.05) |
| Fixed Effects: | | | | | |
| Minimum education level | ✓ | ✓ | ✓ | ✓ | ✓ |
| Year | ✓ | ✓ | ✓ | ✓ | ✓ |
| Region | ✓ | ✓ | ✓ | ✓ | ✓ |
| Industry | ✓ | ✓ | ✓ | ✓ | ✓ |
| Occupation | ✓ | ✓ | ✓ | ✓ | ✓ |
| R-squared | 0.67 | 0.67 | 0.67 | 0.67 | 0.68 |
| R-squared Adj. | 0.67 | 0.67 | 0.67 | 0.67 | 0.67 |
| N | 9990 | 9990 | 9990 | 9990 | 9990 |

Standard errors in parentheses. * $p<0.1$, ** $p<0.05$, *** $p<0.01$

**Table S12. Regression Results for Wage Effects for “Substitutable” Skills.** Results from linear regressions relating the requirement for “substitutable” skills to wages offered within AI-related roles.

### S4.C. External Effects Results

This section presents additional results from the external effects analysis, which investigates the relationship between AI prevalence and skills demanded in non-AI roles. This material provides further details regarding the results displayed in Figure 3 in the main manuscript. Figures S6 and S7 plot the relationship between AI prevalence and “complementary” or “substitutable” skills in non-AI jobs at the company, industry, and region level, for the United Kingdom, and Australia and New Zealand. As in Figure 3, the graphs reveal changing trends between the earlier and later years of our sample (2018-2019 versus 2023-2024). A positive relationship between AI prevalence and demand for “complementary” skills in non-AI roles can be observed at the company, industry, region levels both for the United Kingdom, and Australia and New Zealand. With the exception of the United Kingdom’s industry level, this relationship is stronger in the later years (2023-2024) of the sample. For “substitutable” skills, demand in non-AI roles varies between slightly increasing, to decreasing, to no significant relationship, depending on the level, geography, and year of the analysis.

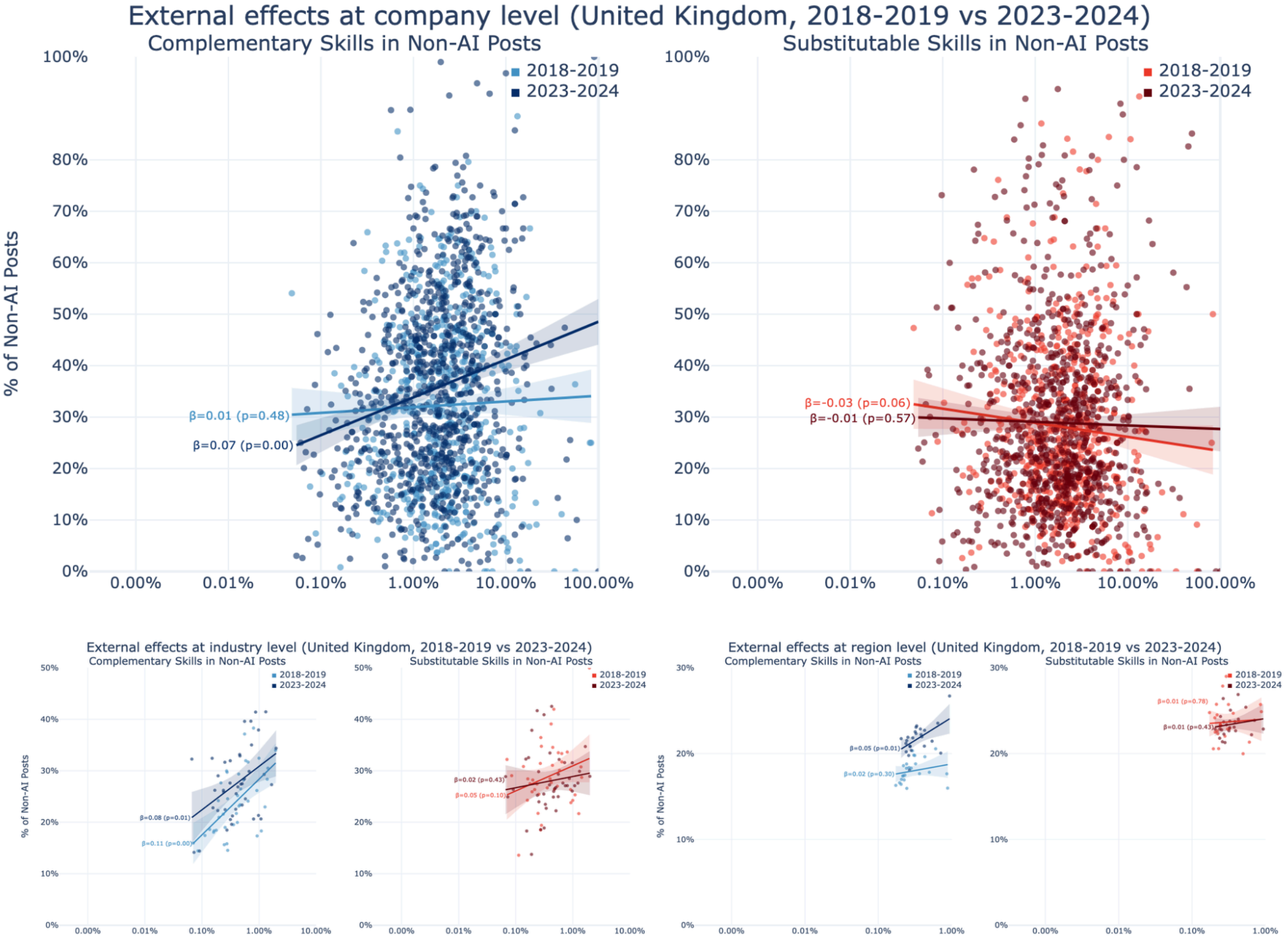

**Figure S7. Descriptive Evidence of External Effects for the United Kingdom, 2018-2024 at the Company, Industry, and Region Level.** The scatterplots reveal the relationship between AI prevalence (the share of AI-related roles) within an entity, and the demand for “complementary” or “substitutable” skills in non-AI roles. For “complementary” skills, a positive relationship can be observed, which appears stronger in the later years (2023-2024) of the sample. For “substitutable” skills, a slight negative relationship can only be read at the company level.

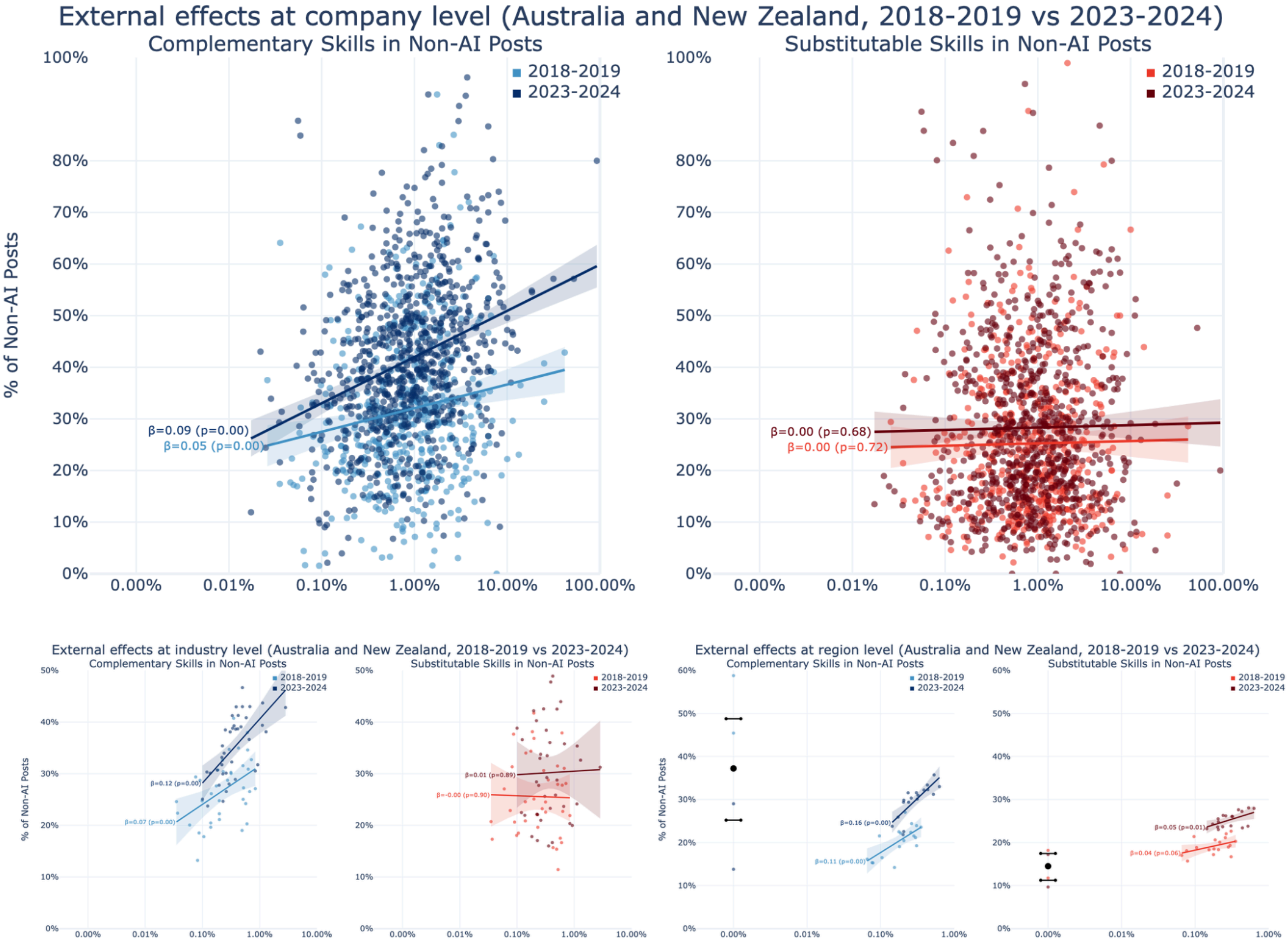

**Figure S8. Descriptive Evidence of External Effects for Australia and New Zealand, 2018-2024 at the Company, Industry, and Region Level.** The scatterplots reveal the relationship between AI prevalence (the share of AI-related roles) within an entity, and the demand for “complementary” or “substitutable” skills in non-AI roles. For “complementary” skills, a positive relationship can be observed, which appears stronger in the later years (2023-2024) of the sample. For “substitutable” skills, the later years bring an upwards shift in skills demand within non-AI roles.

Building on to the simple evidence depicted in Figures S6, S7, as well as the scatter plots in Figure 3, a fractional logistic regression is employed to capture the effect of AI prevalence on skills demand in non-AI roles, including a set of controls and fixed effects as described in section S3. The complete results, as well as the calculated marginal effects, previously presented in the lower panel of Figure 3, are reported in Table S13.

It is particularly notable that Australia and New Zealand exhibit strong marginal effects of AI prevalence at the regional level. A 1 percentage point increase in AI prevalence within a region is here associated with a 18.23 percentage point increase in the demand for “complementary” skills in non-AI roles. By contrast, the marginal effects in the United States and the United Kingdom are 4.46 and 1.25 percentage points respectively. To understand the data composition leading to this interesting result, we plot the relationship between AI prevalence and demand for skills on a linear axis in Figure S8, with different colours distinguishing between different regions. This aims to reveal how each region contributes to the effect captured by our regression analysis. Among inspection of the graphs, the positive relationship between AI prevalence and “complementary” skills in non-AI jobs is particularly pronounced for Australia and New Zealand. It can further be observed that regions change values over the years and do not stick around the same points, so it appears unlikely that a few select regions are driving the regression results reported and calculated in Table S13.

| | United States | | United Kingdom | | Australia & New Zealand | |
|---|---|---|---|---|---|---|
| **Dependent variables:** | **Complementary Skills** | **Substitutable Skills** | **Complementary Skills** | **Substitutable Skills** | **Complementary Skills** | **Substitutable Skills** |
| **— Company Level —** | | | | | | |
| AI prevalence (share AI-related roles) | 2.35*** | -3.63*** | 3.48*** | -0.85 | 7.23*** | -1.61 |
| | (0.55) | (0.65) | (0.87) | (1.03) | (1.26) | (1.54) |
| Intercept | -0.60*** | -0.13 | -1.37*** | -0.69*** | -0.69*** | -0.39*** |
| | (0.10) | (0.11) | (0.18) | (0.19) | (0.13) | (0.14) |
| AI Marginal Effects (calculated) | 0.57% | -0.81% | 0.75% | -0.17% | 1.66% | -0.31% |
| Fixed Effects: | | | | | | |
| Year | ✓ | ✓ | ✓ | ✓ | ✓ | ✓ |
| Industry | ✓ | ✓ | ✓ | ✓ | ✓ | ✓ |
| N | 6469 | 6469 | 2545 | 2545 | 2243 | 2243 |
| AIC | 6187.94 | 5918.58 | 2276.75 | 2199.23 | 2059.05 | 1873.95 |
| **— Industry Level —** | | | | | | |
| AI prevalence (share AI-related roles) | 47.06*** | -15.89** | 31.98*** | 6.81 | 37.89*** | 3.68 |
| | (4.82) | (6.89) | (5.87) | (5.02) | (6.92) | (11.48) |
| Intercept | -0.16 | -0.44 | -0.68*** | -0.46*** | -0.25 | 0.42 |
| | (0.15) | (0.27) | (0.17) | (0.14) | (0.20) | (0.26) |
| AI Marginal Effects (calculated) | 10.77% | -3.74% | 6.08% | 1.38% | 7.94% | 0.74% |
| Fixed Effects: | | | | | | |
| Year | ✓ | ✓ | ✓ | ✓ | ✓ | ✓ |
| Industry | ✓ | ✓ | ✓ | ✓ | ✓ | ✓ |
| N | 138 | 138 | 140 | 140 | 130 | 130 |
| AIC | 137.08 | 140.35 | 125.85 | 129.99 | 124.12 | 122.15 |
| **— Region Level —** | | | | | | |
| AI prevalence (share AI-related roles) | 20.23*** | -6.65*** | 7.90 | 6.89 | 99.88*** | 23.13** |
| | (2.45) | (2.52) | (5.67) | (4.21) | (10.86) | (9.26) |
| Intercept | -1.03*** | -0.79*** | -1.95*** | -0.10 | -2.28*** | -1.93*** |
| | (0.05) | (0.05) | (0.17) | (0.17) | (0.14) | (0.11) |
| AI Marginal Effects (calculated) | 4.46% | -1.51% | 1.25% | 1.24% | 18.23% | 3.96% |
| Fixed Effects: | | | | | | |
| Year | ✓ | ✓ | ✓ | ✓ | ✓ | ✓ |
| Industry | ✓ | ✓ | ✓ | ✓ | ✓ | ✓ |
| N | 353 | 353 | 83 | 83 | 62 | 62 |
| AIC | 314.01 | 319.81 | 74.61 | 79.49 | 64.17 | 62.39 |

Standard errors in parentheses. * $p<0.1$, ** $p<0.05$, *** $p<0.01$

**Table S13. Regression Results for External Effects Analysis.** Results from fractional logistic regressions capturing how AI prevalence affects demand for “complementary” and “substitutable” skills in non-AI roles within companies, industries, and regions. The analysis is conducted for the United States, the United Kingdom, and Australia and New Zealand. The marginal effects of AI prevalence on skills demand are calculated and reported as well.

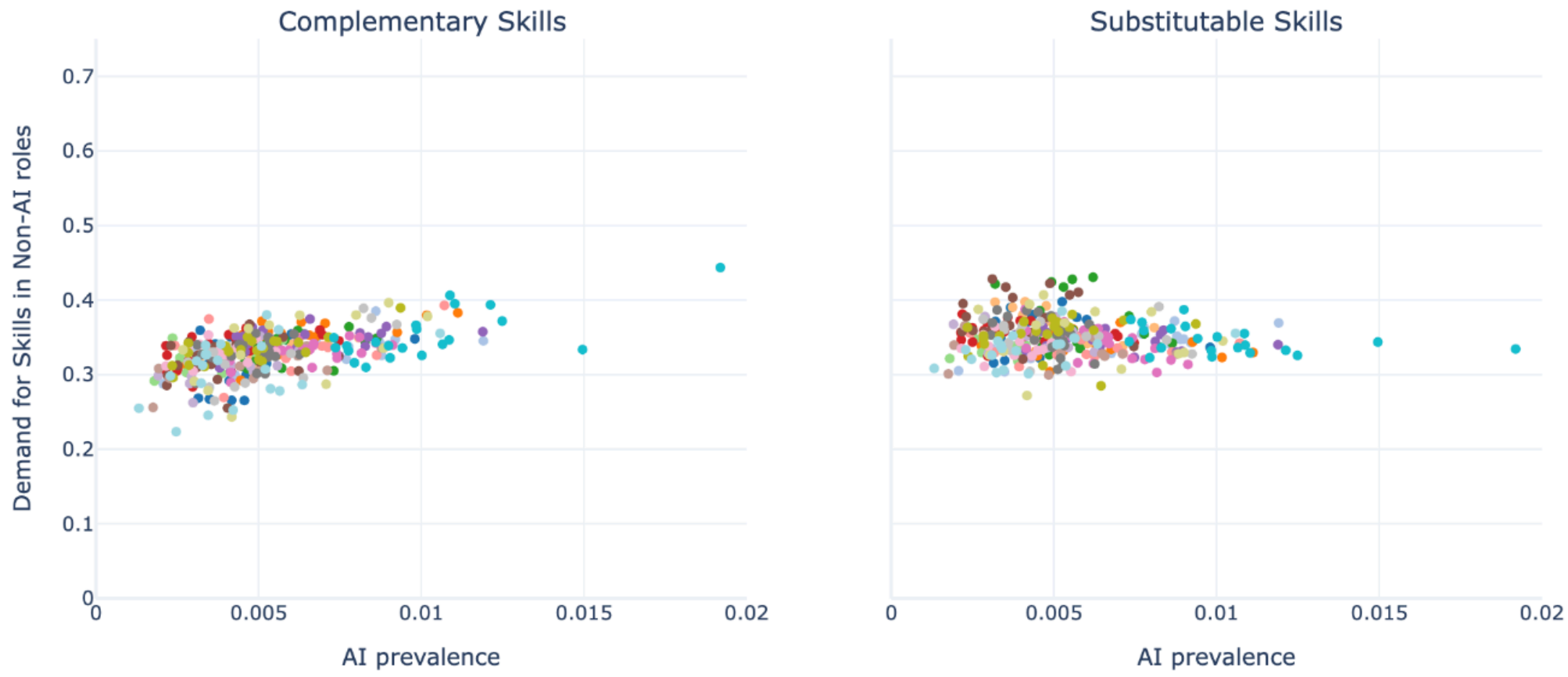

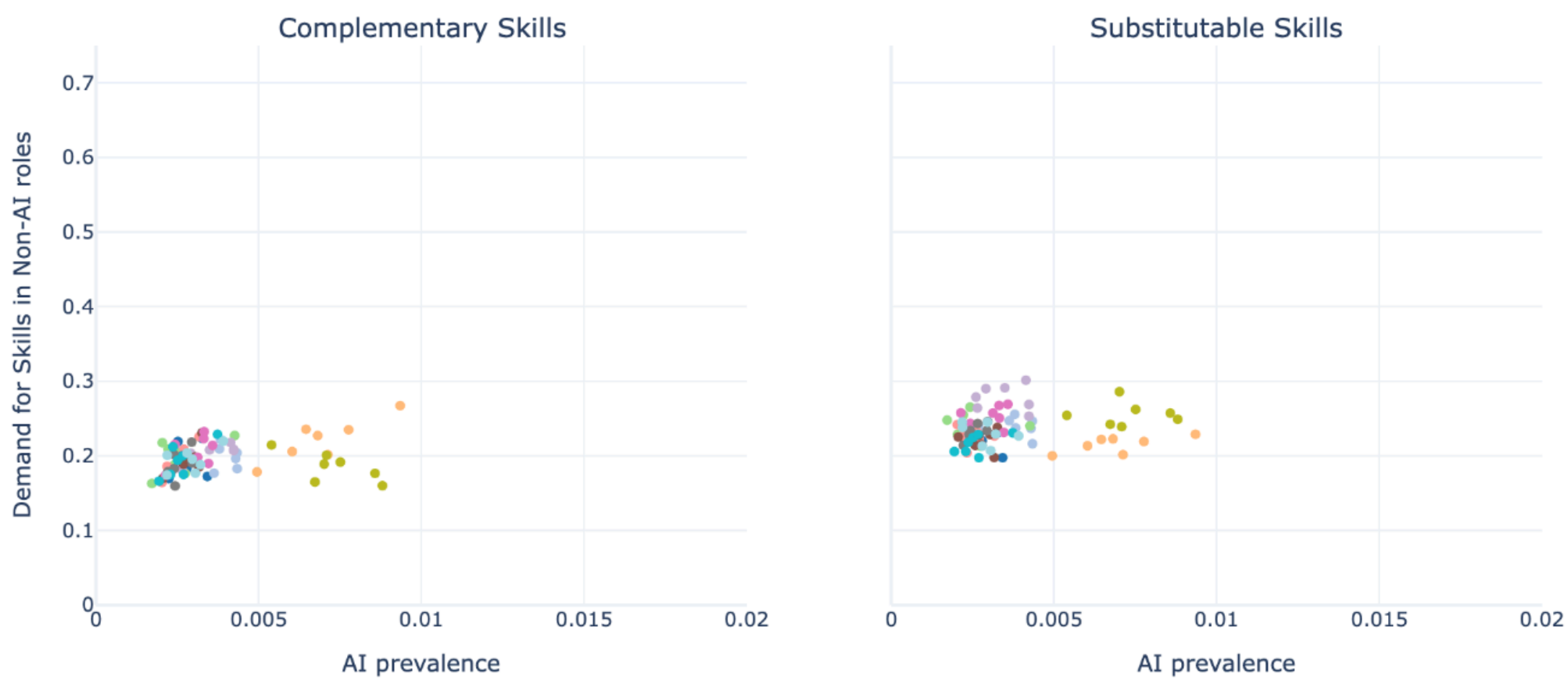

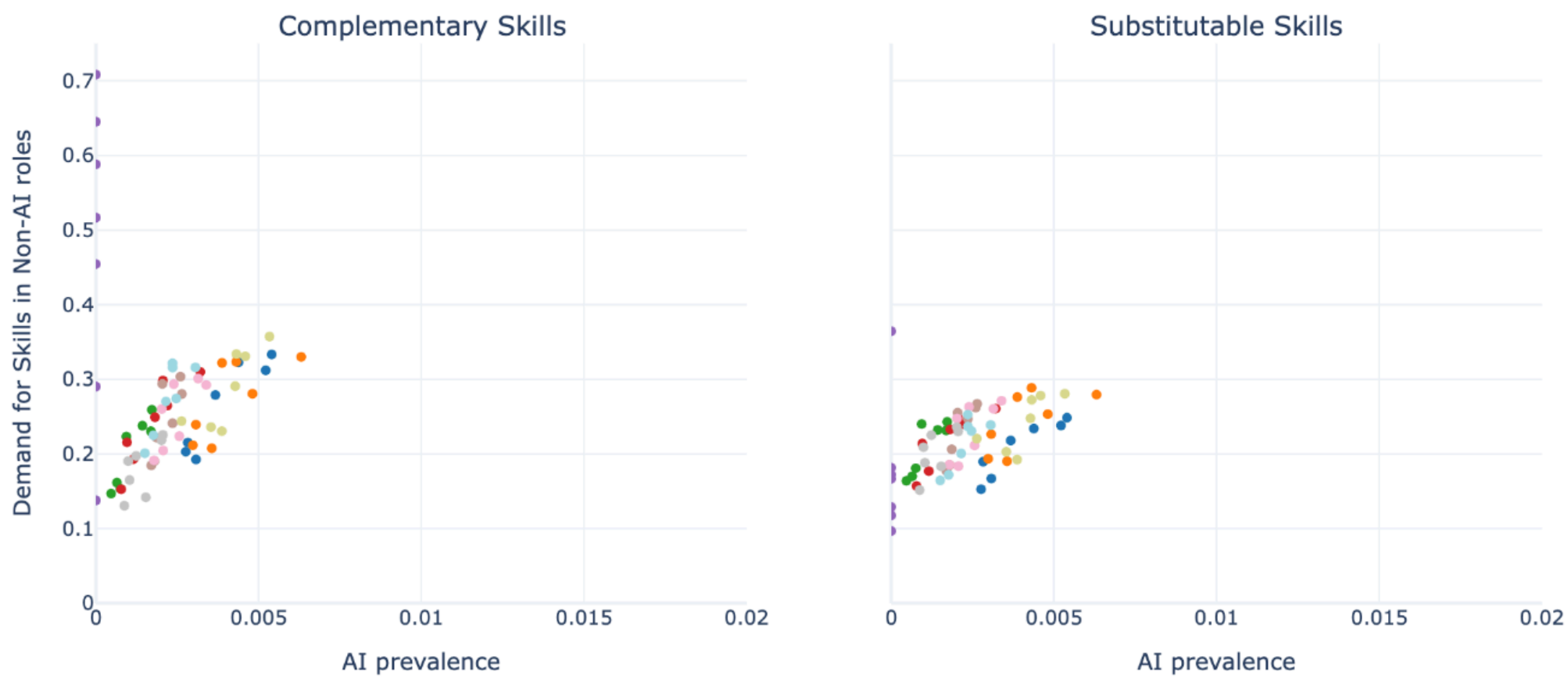

**Figure S9. Skills Demand and AI Prevalence at the Region Level between 2018-2024 for the United States, the United Kingdom, Australia and New Zealand.** The scatter plots reveal the positions of different regions for each geography to understand whether the external effects results are driven by patterns within or across different regions. Each datapoint represents a region-year pair.

We proceed to investigate how the selected skills clusters contribute to the overall effects captured for complementary or substitutable skills. First, we conduct the internal and external analyses outlined in sections S3.A and S3.B on 30 random skills subsamples, each using 75% of skills. Figure S10 visualises the distributions of coefficients resulting from the exercise. To understand how each individual skill contributes to the overall effect documented for 'complementary' or 'substitutable' skill, we measure the impact of any given skill on the coefficients. The results are presented in Figure S11. The external effects of decreased demand for 'substitutable' skills in non-AI roles are largely driven by the decreased demand for 'Customer Service' and 'Customer Support' skills. As AI-related roles become more common within companies, industries, and regions, such customer service related skills are demanded less frequently. A similar pattern can be observed in the internal effects analysis. For the external effects on 'complementary' skills, the positive demand effect is most commonly associated with the 'Problem-Solving' and 'Analytical Thinking' skills.

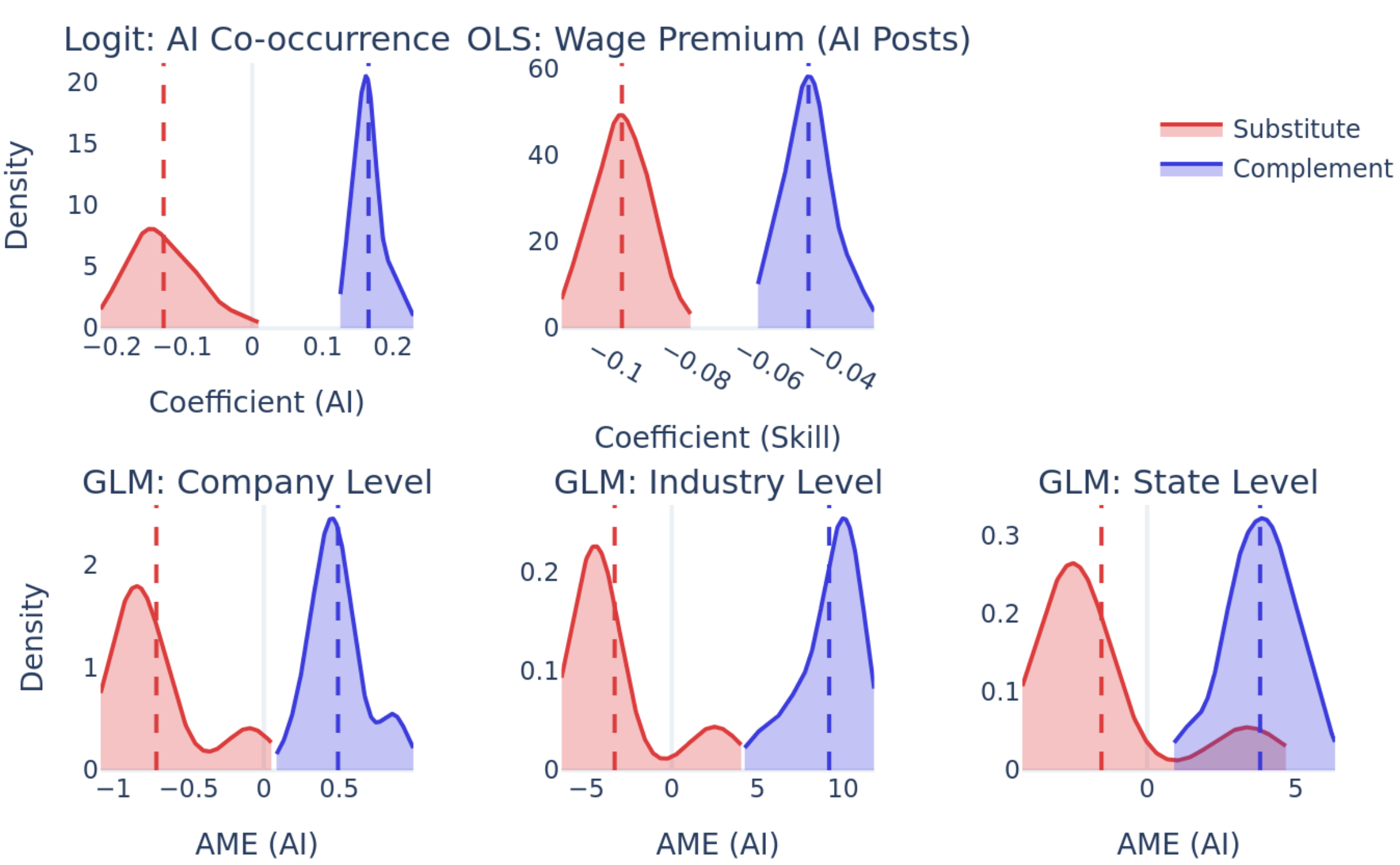

**Figure S10. Density plots reporting the distributions of coefficients found for internal and external effects analyses when randomly varying the skills used to define 'complementary' and 'substitutable' skills.** The upper panel reports coefficients from the internal effects analysis outlined in section S3.A, while the lower panel reports coefficients from the external effects analysis described in section S3.B.

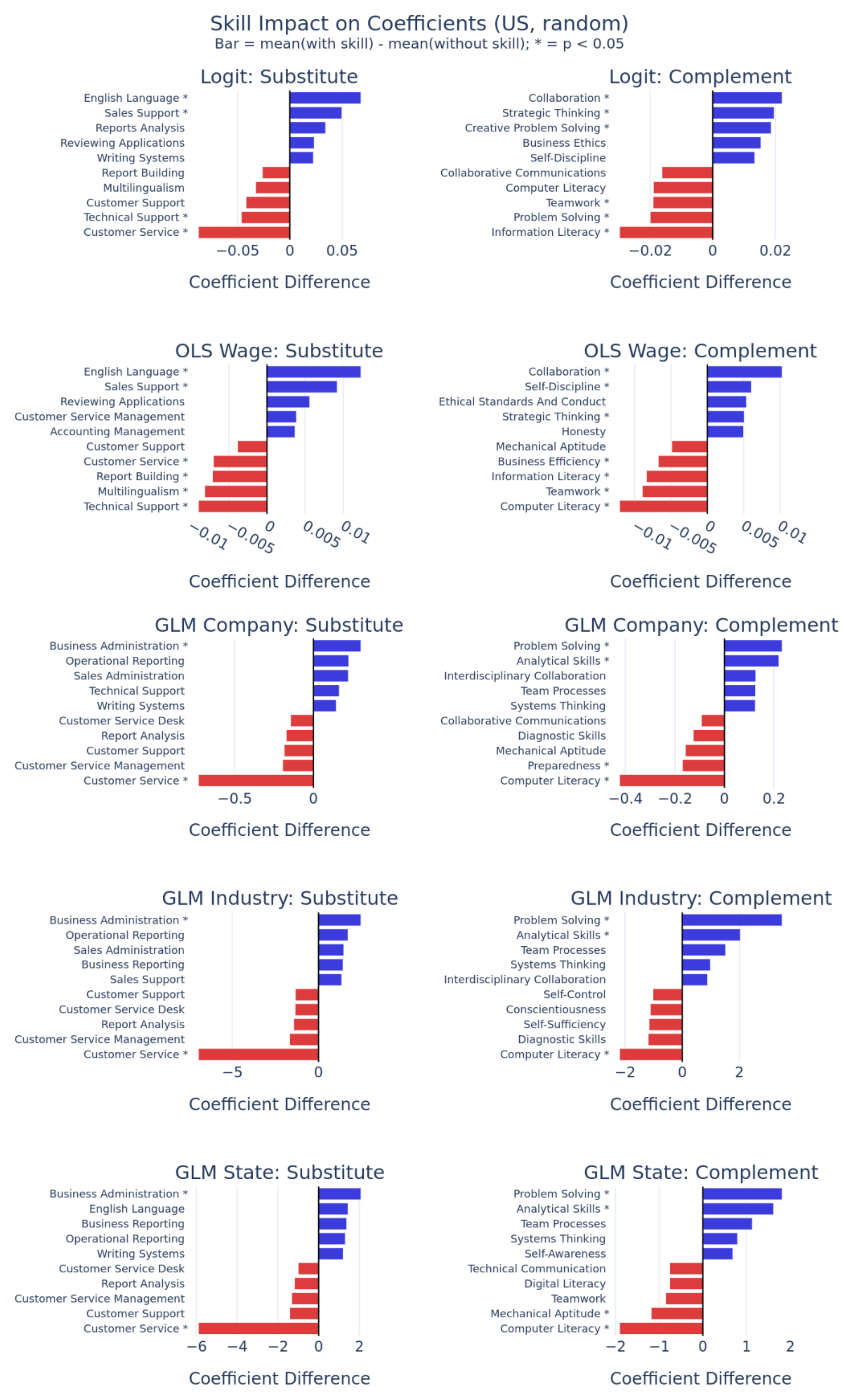

**Figure S11. Contribution of individual skills cluster to the Internal and External Effects Results.** The bar graphs plot the difference in coefficients reported for the analysis with and without each given skill. This exercise reveals how each skills cluster contributes to the overall effect reported for the demand of 'complementary' and 'substitutable' skills.

To verify the robustness of the external effects documented, we repeat our analysis for substitutable skills, omitting “Customer Service” skills, which, as shown in Figure S11, are the main contributors to the decreased demand effect observed. Figure S12 shows that, without “Customer Service” skills, only a very small negative relationship between AI prevalence and substitutable skills can be traced in our data of firms in the United States.

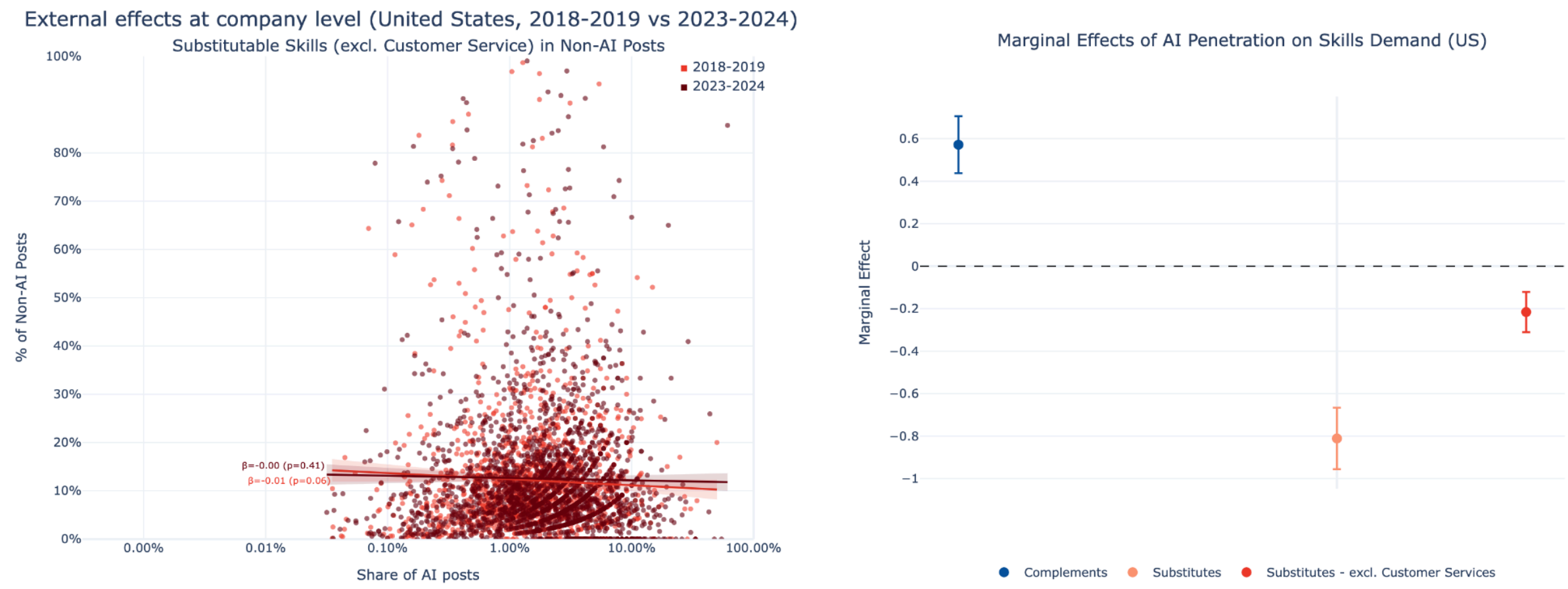

**Figure S12. Robustness check of external effects results, excluding “Customer Service” skills from the set of substitutable skills.** The scatter plot in the left panel relates the share of AI posts within a firm to the percentage of non-AI roles that require substitutable skills. The coefficients plotted in the right panel report a weaker negative relationship between AI prevalence and demand for substitutable skills, when “Customer Service” skills are excluded from the analysis.

We further present the results from the extensive margin analysis, investigating how AI presence, rather than AI prevalence, affects skills demanded in non-AI roles. As outlined in section S3.B, this analysis was performed at the company level, where AI presence is captured by a dummy variable indicating whether the given company has at least one AI-related role. The results reported in Table S14 indicate positive effects on demand for “complementary” skills and negative effects for “substitutable” skills. To illustrate, in the United States, a company with at least one AI-related role is associated with a 5.93 percentage point higher demand for “complementary” skills in its non-AI roles than a company that advertises no AI-related roles. For “substitutable” skills this demand difference lies at -2.64 percentage points.

| | United States | | United Kingdom | | Australia & New Zealand | |
|---|---|---|---|---|---|---|
| **Dependent variables:** | **Complementary Skills** | **Substitutable Skills** | **Complementary Skills** | **Substitutable Skills** | **Complementary Skills** | **Substitutable Skills** |
| | | | — Company Level — | | | |
| AI (yes/no) | 0.25*** | -0.12*** | 3.48*** | -0.85 | 0.28*** | -0.04* |
| | (0.01) | (0.01) | (0.87) | (1.03) | (0.02) | (0.02) |
| Intercept | -0.85*** | -0.39*** | -1.37*** | -0.69*** | -0.81*** | -0.65*** |
| | (0.04) | (0.04) | (0.18) | (0.19) | (0.07) | (0.07) |
| AI Marginal Effects (calculated) | 5.93% | -2.64% | 0.75% | -0.17% | 5.97% | -0.82% |
| Fixed Effects: | | | | | | |
| Year | ✓ | ✓ | ✓ | ✓ | ✓ | ✓ |
| Industry | ✓ | ✓ | ✓ | ✓ | ✓ | ✓ |
| N | 27831 | 27831 | 12956 | 12956 | 9625 | 9625 |
| AIC | 26751.48 | 26519.80 | 10703.74 | 11385.43 | 8558.87 | 8120.49 |

Standard errors in parentheses. * $p<0.1$, ** $p<0.05$, *** $p<0.01$

**Table S14. Regression Results for External Effects Analysis, Extensive Margin.** Results from fractional logistic regressions capturing how the presence of at least 1 AI-related role within a company

affects demand for “complementary” and “substitutable” skills in non-AI roles. The analysis is conducted for the United States, the United Kingdom, and Australia and New Zealand. The marginal effects of AI presence on skills demand are calculated and reported as well.

## S5. Compositional Analysis

The qualitative results of the internal effects analysis are robust to many specifications of the model with various different controls listed in Table S8. The one exception is when we control for the number of skills per post. Adding this control decreases the AI role coefficients for both “substitutable” and “complementary” skill demand. In particular, it makes the “complementary” skills coefficient negative, which changes the qualitative results. However, we find that adding this control always pushes down the relationship between any two groups of skills. The plots in Figure S13 relate “complementary” and “substitutable” skills with each other and a random sample of skills. When including the "SKILL_COUNT" control, the coefficients are always pushed downwards.

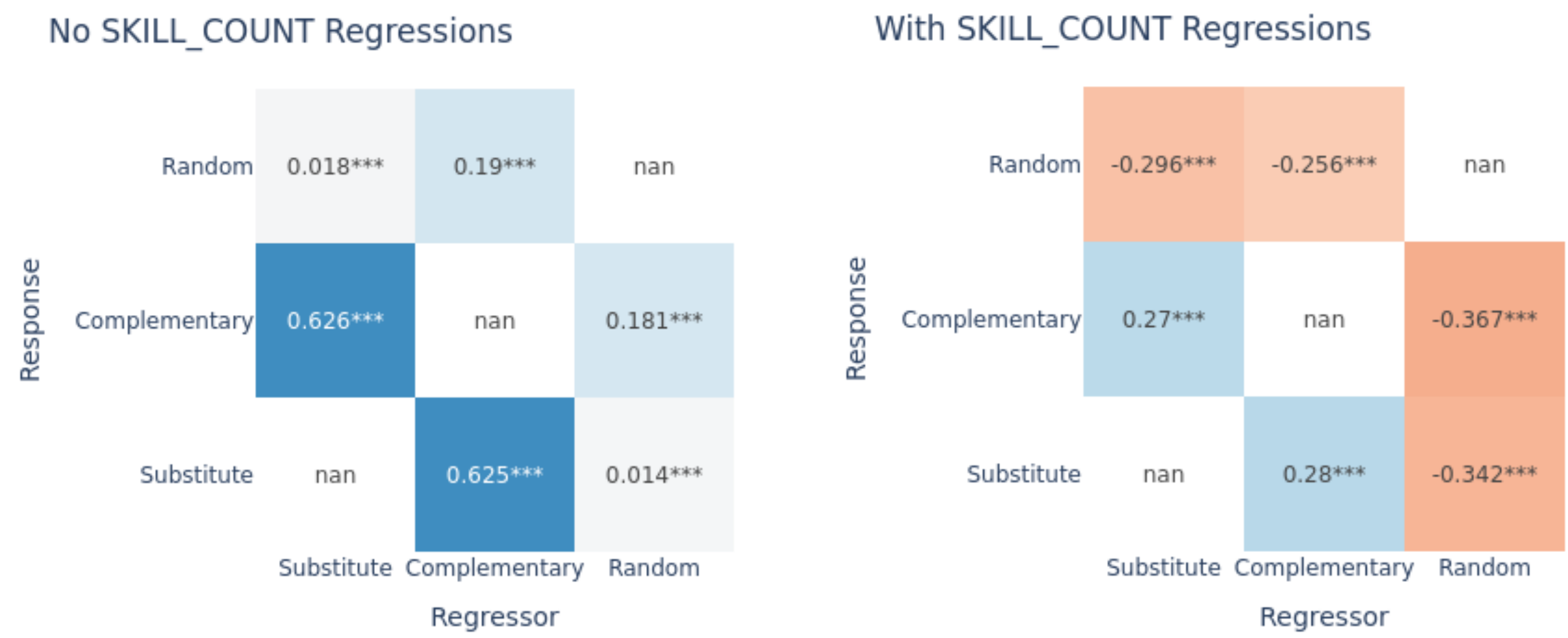

**Figure S13. Relating complementary and substitutable skills with each other and a random sample of skills.** The left panel does not control for the number of skills per job posting while the right panel does. All coefficients decrease when the number of skills is controlled for.

Hence, including the "SKILL_COUNT" control may induce a compositional effect whereby vacancies with the same number of skills must tradeoff which skills they have (e.g. one more AI skill means one less potential “substitutable” skill). In a compositional data structure, out-of-the-box regression methods cannot be applied due to the structural tradeoff in a composition. Therefore, we treat the AI and “complementary” or “substitutable” skills in a compositional analysis. This is useful if we assume that there is some type of skill requirement "budget" that job posts must consider when deciding which skill requirements to include in a post. To add one more AI skill means one less “substitutable” skill. In this setting, the presence of any distinct set of skills will tend to be negatively correlated with each other due to this tradeoff effect. The ternary plot in Figure S14 captures the distribution of AI job postings across the skills compositions. Apart from the majority of skills posted falling into the “other” category, the plots present descriptive evidence that “complementary” skills tend to coincide with AI skills more than “substitutable” skills do.

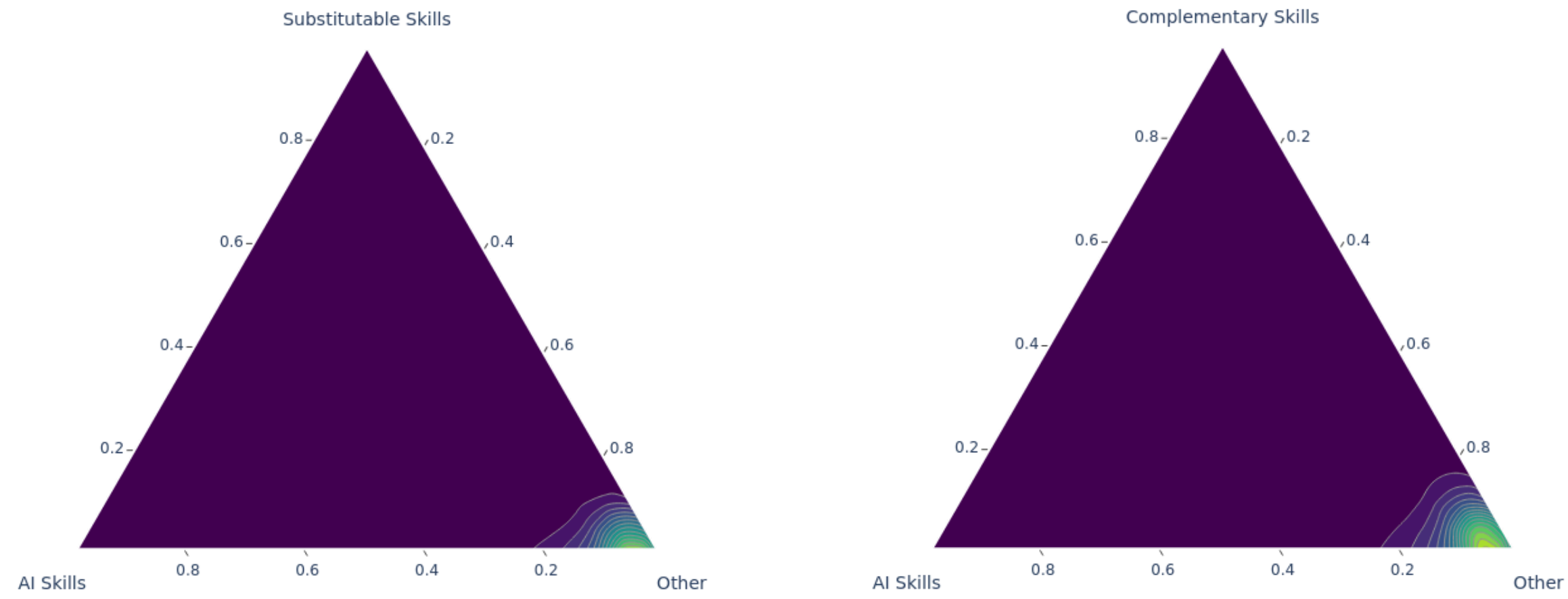

**Fig. S14. Ternary plots mapping the distribution of job postings across skills compositions.** The skills category "Other" refers to the complement of AI skills + "substitutable" skills, or AI skills + "complementary" skills, respectively. "Complementary" skills appear more likely to coincide with AI skills.

### SI References